\documentclass[11pt,a4paper]{article}
\usepackage[english]{babel}
\usepackage{amsmath,amsthm,amssymb,epsfig,latexsym}
\usepackage{color}
\usepackage{ulem}

\newcommand{\so}{\scriptscriptstyle \rm I}
\newcommand{\st}{\scriptscriptstyle \rm I\hspace{-1pt}I}
\newcommand{\sth}{\scriptscriptstyle \rm I\hspace{-1pt}I\hspace{-1pt}I}

\newcommand{\si}{\rm i}
\newcommand{\sii}{\rm ii}
\newcommand{\siii}{\rm iii}
\newcommand{\sv}{\rm v}
\newcommand{\siv}{\rm iv}

\newcommand{\bu}{\bar u}
\newcommand{\bv}{\bar v}
\newcommand{\bx}{\bar x}

\newcommand{\bt}[2]{\bar t^{\{#1\};#2}}

\newcommand{\bw}{\bar w}

\newcommand{\be}[1]{\begin{equation}\label{#1}}
\newcommand{\ba}[1]{\begin{multline}\label{#1}}
\newcommand{\ee}{\end{equation}}
\newcommand{\ea}{\end{multline}}

\newtheorem{prop}{Proposition}[section]
\newtheorem{lemma}{Lemma}[section]
\newtheorem{cor}{Corollary}[section]
\newtheorem{Def}{Definition}[section]

\def\qed{\hfill\nobreak\hbox{$\square$}\par\medbreak}
 \makeatletter
 \@addtoreset{equation}{section}
 \makeatother
 
\newcommand{\bea}{\begin{eqnarray}}
\newcommand{\eea}{\end{eqnarray}}

\def\BB{{\mathbb{B}}}
\def\CC{{\mathbb{C}}}
\def\hBB{{\hat{\mathbb{B}}}}
\def\hCC{{\hat{\mathbb{C}}}}
\newcommand{\ZZ}{{\mathbb Z}}

\def\rvac{|0\rangle}
\def\lvac{\langle 0 |}

\def\EE{{\rm E}}

\def\r#1{(\ref{#1})}
\def\sk#1{\left(#1\right)}
\def\Dfun{\Delta}

\def\prt#1{[#1]}
\def\cci#1{c_{\prt{#1}}}
\def\Fli#1{\gamma_{#1}}
\def\hFli#1{\hat\gamma_{#1}}

\newcommand{\ort}{{\scriptscriptstyle\overrightarrow{\displaystyle t}}}
\newcommand{\olt}{{\scriptscriptstyle\overleftarrow{\displaystyle t}}}

\newcommand{\olx}{{\scriptscriptstyle\overleftarrow{\displaystyle x}}}

\begin{document}

\thispagestyle{empty}
\setcounter{page}{0}

\vspace{12pt}

\begin{center}
\begin{LARGE}
{\bf Actions of the monodromy matrix elements\\[1ex]
 onto $\mathfrak{gl}(m|n)$-invariant Bethe vectors}
\end{LARGE}

\vspace{40pt}

\begin{large}
{A.~Hutsalyuk${}^{a}$, A.~Liashyk${}^{b}$,
S.~Z.~Pakuliak${}^{c,d}$,\\ E.~Ragoucy${}^f$, N.~A.~Slavnov${}^{g}$\  \footnote{
hutsalyuk@gmail.com, a.liashyk@gmail.com, stanislav.pakuliak@jinr.ru, eric.ragoucy@lapth.cnrs.fr, nslavnov@mi-ras.ru}}
\end{large}

\vspace{10mm}

${}^a$ {\it BME “Momentum” Statistical Field Theory Research Group,
Department of Theoretical Physics,
Budapest University of Technology and Economics,
1521 Budapest, Hungary}

\vspace{2mm}

${}^b$ {\it Skolkovo Institute of Science and Technology, Moscow, Russia}

\vspace{2mm}

${}^c$ {\it Moscow Institute of Physics and Technology,  Dolgoprudny, Moscow reg., Russia}

\vspace{2mm}

${}^d$ {\it Laboratory of Theoretical Physics, JINR,  Dubna, Moscow reg., Russia}

\vspace{2mm}

${}^f$ {\it Laboratoire de Physique Th\'eorique LAPTh, CNRS and USMB,\\
BP 110, 74941 Annecy-le-Vieux Cedex, France}

\vspace{2mm}

${}^g$ {\it Steklov Mathematical Institute of Russian Academy of Sciences,\\ Moscow, Russia}

\end{center}


\vspace{4mm}

\begin{abstract}
Multiple  actions of the monodromy matrix elements onto off-shell Bethe vectors in the
$\mathfrak{gl}(m|n)$-invariant quantum integrable models are calculated.
These actions are used to describe recursions for
the highest coefficients in the sum formula for the scalar product.
For simplicity, detailed proofs are given for the  $\mathfrak{gl}(m)$ case. The results for the
supersymmetric case can be obtained similarly and are formulated without proofs.
\end{abstract}

\newpage

\section{Introduction}

This paper is a continuation of the paper \cite{HLPRS17} devoted to the description
of the off-shell Bethe for the $\mathfrak{gl}(m|n)$-invariant quantum integrable models. One
of the main results of  \cite{HLPRS17}  was an action formula of the
 upper-triangular and diagonal monodromy
matrix elements onto off-shell Bethe vectors in the corresponding models.
These results were obtained  by expressing the Bethe vectors
in terms of the current generators of the Yangian double $DY(\mathfrak{gl}(m|n))$.
The same method of calculation for the actions of the lower-triangular monodromy
matrix elements appears to be too cumbersome to be detailed. The present paper describes
an alternative way to find these actions using a generalization of the so-called {\it zero modes method}.
To simplify our presentation we give the detailed proofs only for the
non-supersymmetric case $n=0$, since the methods we are using extend readily
 to the general case.

The action of the monodromy matrix elements onto the Bethe vectors plays important role in the study of quantum integrable models. The action of the upper-triangular elements generate recursions for the Bethe vectors. Formulas for the action of the diagonal elements are key in solving the problem of the spectrum of Hamiltonians of quantum integrable systems. It is from these formulas that the Bethe equations that determine the spectrum follow. Finally, the action of the lower-triangular elements are necessary for the studying scalar products of Bethe vectors, which, in their turn, are used for calculating correlation functions. In this case, we need formulas for the action of not only a single element of the monodromy matrix, but also the so-called multiple action formulas when we act on the Bethe vector by a product of lower-triangular elements.

Let $N=m+n-1$. First, we focus on calculating the action of monodromy matrix elements onto off-shell Bethe vectors for $\mathfrak{gl}(N+1)$-invariant integrable models.
Recall that the paper  \cite{HLPRS17} discusses an approach to a description of the space of states for the quantum integrable models using infinite-dimensional
current algebras  proposed in \cite{EKhP07} and developed in \cite{KhP-Kyoto}. This approach takes advantage of the fact that monodromy matrices in
the quantum integrable models satisfy the same commutation relations as a generating series of the generators of certain infinite-dimensional algebras \cite{D88}.
These algebras can usually be realized in two different patterns, either in the form of so-called $L$-operators or in terms of total currents \cite{DF93,JLM18}.
The description of the space of states (Bethe vectors) in quantum integrable models uses the concept of projections onto intersections of the different
type Borel subalgebras related either to $L$-operator or current realizations respectively.

The fact that the action of monodromy
matrix elements onto Bethe vectors produces a linear combination of the same vectors
is almost obvious within the projection method. However, obtaining explicit and effective formulas for this action  is a rather complex combinatorial problem.
In the paper \cite{HLPRS17}, only the actions of the upper-triangular and diagonal elements were calculated. In this paper we
present an alternative method to find the actions of all monodromy matrix elements. For this, we
use only information about the action of the element $T_{1,N+1}(z)$ and the zero mode operators
$T_{i+1,i}[0]$ onto off-shell Bethe vectors and commutation relations between them.
This starting  information can be easily obtained from the projection method and it is formulated
as lemma~\ref{prop2} below.

We call the method of calculating the monodromy matrix elements action onto Bethe vector
the {\it zero modes method}, because it is based on the commutation
relations
\begin{equation}\label{zm}
[T_{i,j}(z),T_{\ell+1,\ell}[0]]=\delta_{i,\ell}\ \kappa_i\ T_{i+1,j}(z)  - \delta_{\ell,j-1}\ \kappa_j\   T_{i,j-1}(z),
\end{equation}
which follows from the basic commutation relations \r{rrt2} described below.

In this paper we do not use the projection method to describe the off-shell Bethe vectors.
We fix these objects by the explicit formulas for the action of the transfer matrix (trace of the
monodromy matrix) onto Bethe vectors and requirement that they become eigenvectors of the
transfer matrix if the parameters of the Bethe vectors satisfy the so-called Bethe equations.

The paper is organized as follows. In section~\ref{Not}, we introduce our notation and describe the commutation relations of
the monodromy matrix elements. Then, in section~\ref{BV}, we define the Bethe vectors,
the dual Bethe vectors and describe their normalization.   Section~\ref{mainres} contains the main
result of the paper. This result for the simplest case of the action of one monodromy matrix
element is proved in  appendix~\ref{ApA}. The general case is proved in
appendix~\ref{ApB}. Section~\ref{spsect} contains applications of the results obtained. Here we formulate recursions for the
highest coefficients of the scalar product of Bethe vectors with respect to the rank of the algebra.
Section~\ref{super} contains a generalization of the above results to the case of $\mathfrak{gl}(m|n)$-integrable models without detailed  proofs.

\section{RTT-algebra and notation}\label{Not}

The quantum integrable models we are dealing with  are treated by the so-called nested algebraic Bethe
ansatz \cite{KulRes81,KulRes83} and correspond to  algebras of rank more than 1.
All these models are described by the operators gathered in the monodromy matrix
 $T(z)$ which acts in a Hilbert space $\mathcal{H}$ and an auxiliary space $\mathbb{C}^{N+1}$. It satisfies an $RTT$ commutation relation
\begin{equation}\label{rtt}
  R(u,v) \left( T(u)\otimes\mathbf{I} \right) \left( \mathbf{I}\otimes T(v) \right) = \left( \mathbf{I}\otimes T(v) \right) \left( T(u)\otimes\mathbf{I} \right) R(u,v).
\end{equation}
Here $\mathbf{I}$ is the identity matrix in $\mathbb{C}^{N+1}$, and $\mathbf{P}$ is a permutation matrix in $\mathbb{C}^{N+1}\otimes \mathbb{C}^{N+1}$.
A $\mathfrak{gl}(N+1)$-invariant\footnote{%
Here we consider
only models corresponding to the algebra $\mathfrak{gl}(N+1)$. Results
for the supersymmetric models are collected in the section~\ref{super}. Similar results
for the integrable models related to other series algebras will be considered elsewhere (see also pioneering
papers \cite{LPRS-19,GR-19}).} $R$-matrix $R(u,v)$  acts in $\mathbb{C}^{N+1}\otimes \mathbb{C}^{N+1}$ and is given by
\begin{equation}\label{Rmat}
  R(u,v) = \mathbf{I}\otimes\mathbf{I} + g(u,v) \mathbf{P}, \quad g(u,v) = \frac{c}{u-v},
\end{equation}
where $c$ is a complex constant. Starting from \r{rtt} one can easily obtain commutation relations for the monodromy matrix elements
\begin{equation}\label{mmel}
T(u)=\sum_{i,j=1}^{N+1}\EE_{ij}\otimes T_{i,j}(u)
\end{equation}
in the form
\begin{equation}\label{rrt2}
  \left[ T_{i,j}(u), T_{k,l}(v) \right] = g(u,v) \left( T_{i,l}(u)T_{k,j}(v) - T_{i,l}(v)T_{k,j}(u) \right),
\end{equation}
where $\EE_{ij}$ is a unit matrix with the only non-zero element equal to 1 on the intersection of the $i$-th
row and $j$-th column.

For the reasons which will become clear later we consider an asymptotic expansion of the
monodromy matrix
\begin{equation}\label{depen}
T_{i,j}(u)=\delta_{ij}\kappa_i+\sum_{\ell\geq0} T_{i,j}[\ell](u/c)^{-\ell-1} ,
\end{equation}
which includes parameters $\kappa_i\in\CC$, $i=1,\ldots,N+1$, in the zeroth order of the expansion. When
all $\kappa_i$ are equal to 1 the expansion \r{depen} corresponds to Yangian\footnote{According to \cite{Mol07}
we define a Yangian as a Hopf algebra generated by coefficients $T_{i,j}[\ell]$,  $\ell \ge 0$,
 such that commutation
relations \eqref{rtt} are satisfied and the asymptotics of $T_{i,j}(u)$ is $T_{i,j}(u)=\delta_{ij}+O(u^{-1})$ at $u\to\infty$.}
$Y(\mathfrak{gl}(N+1))$ \cite{D88}. Using commutativity of the $R$-matrix \r{Rmat}
with $\mathbb{K}\otimes\mathbb{K}$, where $\mathbb{K}={\rm diag}(\kappa_1,\ldots,\kappa_{N+1})$
is a diagonal matrix, we can multiply the Yangian $RTT$ commutation relations \eqref{rtt}
by $\mathbb{K}\otimes\mathbb{K}$ to obtain this expansion of the  monodromy matrix.

Commutation relations \r{rtt} imply that the transfer matrices
\begin{equation}\label{trans}
\mathfrak{t}(z)=\sum_{i=1}^{N+1}T_{i,i}(z)
\end{equation}
commute for arbitrary values of the spectral parameters
\begin{equation*}
\mathfrak{t}(u)\cdot \mathfrak{t}(v)=\mathfrak{t}(v)\cdot \mathfrak{t}(u).
\end{equation*}
Thus,  $\mathfrak{t}(z)$ generates a set (in general infinite) of commuting quantities. Solving
the model by the algebraic Bethe ansatz amounts to find  eigenvectors of
the transfer matrix $\mathfrak{t}(z)$ in the Hilbert space $\mathcal{H}$ of the
quantum integrable model. To solve this problem the Hilbert space $\mathcal{H}$ should
possess a special vector $\rvac$ called reference state such that
\begin{equation}\label{tii}
\begin{aligned}
   T_{i,i}(u)\rvac  &=  \lambda_{i}(u) \rvac, \quad & & i=1,...,N+1,\\
  T_{i,j}(u)\rvac  &=  0,  & &i>j.
\end{aligned}
\end{equation}
Here the functional parameters $\lambda_{i}(u)$ are characteristic of the concrete model.
Further on we will use the ratios of these free functional parameters
\begin{equation}\label{ratios}
  \alpha_i(u) = \frac{\lambda_{i}(u)}{\lambda_{i+1}(u)},\quad
 i=1, \ldots, N,
\end{equation}
with the asymptotic values
\begin{equation*}
 \lim_{u\to\infty}\alpha_i(u) = \frac{\kappa_{i}}{\kappa_{i+1}}, \quad
   i=1, \ldots, N .
\end{equation*}

\subsection{Notation}

Besides rational function $g(u,v)$ already introduced by \r{Rmat}, we  define two
rational functions $f(u,v)$ and $h(u,v)$ by
\begin{equation}\label{f}
  f(u,v) = 1 + g(u,v) = \frac{u-v+c}{u-v}, \quad  h(u,v) =\frac{ f(u,v)}{ g(u,v)} =  \frac{u-v+c}{c}.
\end{equation}

For any of  the functions $a=g,f,h$  and
for any set of  generic complex parameters  $\bar x=\{x_1,\ldots,x_p\}$
we introduce `triangular' products
\begin{equation}\label{tr-pr}
 \Dfun_a(\bar x)=\prod_{i<j}^pa(x_j,x_i) ,\qquad
\Dfun'_a(\bar x)=\prod_{i<j}^pa(x_i,x_j).
\end{equation}
 For any two sets of  generic complex parameters  $\bar y=\{y_1,\ldots,y_p\}$ and $\bar x=\{x_1,\ldots,x_p\}$
of the same cardinality $\#\bar y=\#\bar x=p$
we also introduce an
Izergin determinant $K(\bar y|\bar x)$
\begin{equation}\label{IzD}
K(\bar y|\bar x)=\Dfun_g(\bar y)\Dfun'_g(\bar x) \prod_{\ell,\ell'=1}^p h(y_\ell,x_{\ell'}) \det\left[
\frac{g(y_\ell,x_{\ell'})}{h(y_\ell,x_{\ell'})}\right]_{\ell,\ell'=1,\ldots,p}\;.
\end{equation}
Any Izergin determinant depending on $\bar x$ and $\bar y$, such that $\#\bar x \ne \#\bar y$, is by definition considered to be equal to zero.
If  $\#\bar x = \#\bar y=0$, then $K(\varnothing|\varnothing)\equiv1$.

Further on we will often work with sets of parameters.
We denote them by a bar, like in \r{tr-pr}, \r{IzD}. In particular, the Bethe vector
$\BB(\bar t)$ depends on the set of parameters (usually called Bethe parameters)
\begin{equation}\label{Bpar}
\bar t=\{\bar t^1,\bar t^2,\ldots,\bar t^N\},\qquad \bar t^i=\{t^i_1,t^i_2,\ldots,t^i_{r_i}\},\qquad
i=1,\ldots,N.
\end{equation}
The upper index labels the type of Bethe parameter and corresponds to the
simple roots of the algebra $\mathfrak{gl}(N+1)$.
To simplify further formulas for the products over sets, we use the following convention:
\begin{equation}\label{SH-prod}
   f(u,\bar t^i)=\prod_{t^i_j\in\bar t^i} f(u, t^i_j), \qquad
   f(\bar t^s,\bar x^p)= \prod_{t^s_j\in\bar t^s}\prod_{x^p_k\in\bar x^p} f(t^s_j,x^p_k),
\end{equation}
\begin{equation}\label{shpr} 
  \lambda_i(\bar t^i)=\prod_{ t^i_j\in\bar t^i} \lambda_i( t^i_j),  \quad\quad
  \alpha_{i}(\bar t^i) = \prod_{ t^i_j\in\bar t^i} \alpha_{i}(t^i_{j}), \quad\quad
  T_{i,j}(\bar t^s_{\so})= \prod_{ t^s_k\in\bar t^s_{\so}} T_{i,j}( t^s_k).
\end{equation}
In a word, if any scalar function or mutually commuting operators\footnote{It follows from \r{rrt2} that $[T_{i,j}(u),T_{i,j}(v)]=0$.} depend on a set of parameters,
we assume  the product
of these quantities with respect to this set. We always assume that any such
product is 1 if any of the sets is empty.

For any set of Bethe parameters $\bar t^i$ of  cardinality $\#\bar t^i=r_i$,
the set $\bar t^i_\ell$ means the set $\bar t^i\setminus \{t^i_\ell\}$ of  cardinality
$r_i-1$.

\section{Bethe vectors}\label{BV}

The Bethe vectors $\BB(\bar t)\in\mathcal{H}$ are rather special polynomials in the non-commuting operators
$T_{i,j}(t)$ for $i\leq j$ acting on the reference vector $\rvac$. We do not use the explicit form of these polynomials,
however, the reader can find it in \cite{HLPRS17}. The main property of Bethe vectors is that they become eigenvectors of the transfer matrix
\begin{equation}\label{BVcon}
\mathfrak{t}(z)\BB(\bar t)=\tau(z;\bar t)\BB(\bar t),
\end{equation}
 provided the Bethe parameters $\bar t$ satisfy a system of equations
\begin{equation}\label{BE}
 \alpha_i(t^i_\ell)=\frac{f(t^i_\ell,\bar t^i_\ell)}{f(\bar t^i_\ell,t^i_\ell)}\ \frac{f(\bar t^{i+1},t^i_\ell)}{f(t^i_\ell,\bar t^{i-1})},
 \qquad  \bar t^0=\bar t^{N+1} =  \varnothing,
 \end{equation}
called the Bethe equations. Then the vector $\BB(\bar t)$ is called {\it on-shell Bethe vector}. Otherwise, if the parameters $\bar t$ are generic
complex numbers, the vector $\BB(\bar t)$ is called {\it off-shell Bethe vector}.
The eigenvalue $\tau(z;\bar t)$ in \eqref{BVcon} is
\begin{equation}\label{BVeig}
\tau(z;\bar t)=\sum_{i=1}^{N+1}\lambda_i(z)f(z,\bar t^{i-1})f(\bar t^i,z).
\end{equation}
It is shown in appendix~\ref{ApC} that the action of the transfer matrix computed via the proposition~\ref{prop1} results in the relation \r{BVcon}
provided the Bethe equations \r{BE} are fulfilled.

 Among all terms in the polynomials defining Bethe vectors, we single out one, which is called the {\it main term}.
Its distinctive property is that it contains only the operators $T_{i,i+1}$ and does not contain the operators $T_{i,j}$ with $j-i>1$.
We fix normalization of the Bethe vectors in such a way that the main term $\widetilde{\mathbb{B}}(\bar t)$ has the form
\begin{equation}\label{mtb}
 \widetilde{\mathbb{B}}(\bar t)=\frac{T_{N,N+1}(\bar t^{N})T_{N-1,N}(\bar t^{N-1}) \cdots
 T_{23}(\bar t^2) T_{12}(\bar t^1) |0\rangle}
	  {\prod_{i=1}^{N}\lambda_{i+1}(\bar t^{i})\prod_{i=1}^{N-1}  f(\bar t^{i+1},\bar t^i)}.
\end{equation}

In order to define the scalar product of the Bethe vectors we need first to define the left (dual)
off-shell Bethe vector. This can be done using transposition antimorphism of the algebra \r{rrt2}
\begin{equation}\label{sp1}
\Psi\ :\ T_{i,j}(u)\to T_{j,i}(u),\qquad \Psi(A\cdot B)=\Psi(B)\cdot \Psi(A),
\end{equation}
where $A$ and $B$ are any products of the monodromy matrix elements. It is easy to see that $\Psi$ is an involution: $\Psi^2={\rm id}$.
We extend this antimorphism to the Hilbert space of the quantum integrable models by the
rule
\begin{equation}\label{sp2}
\Psi(\rvac)=\lvac,\qquad \Psi(A\rvac)=\lvac\Psi(A),
\end{equation}
with normalization $\langle0|0\rangle=1$.
Application of the antimorphism $\Psi$ to the formulas \r{tii} yields
\begin{equation}\label{sp3}
\begin{aligned}
  & \lvac T_{i,i}(u)  =  \lambda_{i}(u) \lvac, \\
  & \lvac T_{j,i}(u)  =  0, \quad i>j.
\end{aligned}
\end{equation}
We define the dual off-shell Bethe vectors $\CC(\bar t)$ as
\begin{equation}\label{sp4}
\CC(\bar t)=\Psi(\BB(\bar t)),
\end{equation}
with normalization
\begin{equation}\label{sp5}
\tilde\CC(\bar t)=\Psi(\tilde\BB(\bar t))=
\frac{ \lvac
 T_{2,1}(\bar t^1) T_{3,2}(\bar t^2)\cdots T_{N,N-1}(\bar t^{N-1}) T_{N+1,N}(\bar t^{N}) }
				      {\prod_{i=1}^{N}\lambda_{i+1}(\bar t^{i})\prod_{i=1}^{N-1}  f(\bar t^{i+1},\bar t^i)}\;.
\end{equation}

\clearpage

\section{Multiple actions\label{mainres}}

The aim of this section is to present the multiple action of the monodromy matrix entry $T_{i,j}(\bar z)$ on Bethe vectors $\BB(\bar t)$. This action will be presented as a sum on some partitions of sets, and to ease the reading, we first describe which type of partitions we will consider.
\begin{Def}\label{def:ij-cond}
Let $\bar z$ be a  set of arbitrary complex variables of cardinality $\#\bar z=p$,  $\bar t=\{\bar t^0, \bar t^1,...,\bar t^{N+1}\}$ be a multi-set with $\bar t^0=\bar t^{N+1}=\emptyset$ and let $i$ and $j$
be arbitrary integers from the set $\{1,...,N+1\}$.
We say that the multi-set $\bar w=\{\bar w^0,\bar w^1,...,\bar w^{N+1}\}$ with $\bar w^s = \{\bar z, \bar t^s\}$ is divided into subsets obeying the $\boldsymbol{(i,j)}$\textbf{-condition w.r.t.} $\boldsymbol{\bar z}$ if it obeys the following conditions:
\begin{itemize}
\item    The sets $\bar w^s$ are divided into three subsets $\{\bar w_{\so}^s,\bar w_{\st}^s, \bar w_{\sth}^s\}\vdash \bar w^s$.
\item   Boundary conditions: $\bar w^0_{\so} = \bar w^{N+1}_{\sth} =
\bar z$, $\bar w^0_{\st}=\bar w^0_{\sth} = \bar w^{N+1}_{\so}=\bar w^{N+1}_{\st} =  \varnothing$.
\item  The subsets $\bar w^s_{\so}$ are non empty only for $s<i$
and have cardinality $\# \bw^s_{\so} = p$ when $s<i$.
\item The subsets $\bar w^s_{\sth}$ are non empty only for $s\ge j$ and have cardinality $\# \bw^s_{\sth} = p$ when $s\ge j$.
\item The rest of the variables belongs to the subsets $\bar w^s_{\st}$.
\end{itemize}
\end{Def}

The calculation of the multiple action relies on knowledge of the (single) action of $T_{i,j}(z)$, which is described in
\begin{lemma}\label{lem1}
The  action of monodromy matrix element $T_{i,j}(z)$ onto the off-shell Bethe vector $\BB(\bar t)$ is given by the expression
\begin{equation}\label{ac222}
\begin{split}
     & T_{i,j}(z) \mathbb{B}(\bar t) =  \lambda_{N+1}(z)
      \sum_{{\rm part}} \mathbb{B}(\bar w_{\st} )
      \frac{\prod_{s=j}^{i-1} f(\bar w^s_{\so},\bar w^s_{\sth})} {\prod_{s=j}^{i-2} f(\bar w^{s+1}_{\so},\bar w^s_{\sth}) }   \\
   &\quad\times    \prod_{s=1}^{i-1} \frac{ f(\bar w^s_{\so},\bar w^s_{\st})}
   {h(\bar w^s_{\so}, \bar w^{s-1}_{\so}) f(\bar w^s_{\so},\bar w^{s-1}_{\st})}
		\prod_{s=j}^{N}      \frac{\alpha_{s}(\bar w^{s}_{\sth})
		 f(\bar w^s_{\st},\bar w^s_{\sth})   }
		{h(\bar w^{s+1}_{\sth}, \bar w^{s}_{\sth})f(\bar w^{s+1}_{\st}, \bar w^{s}_{\sth})}.
		      \end{split}
\end{equation}
 The sum in \r{ac222} runs over partitions obeying the ${(i,j)}$-condition w.r.t. $\{z\}$.
\end{lemma}
The proof of this lemma is given in appendix \ref{ApA}.
It uses the zero mode method based on
the commutation relations  \r{zm}
and the following
\begin{lemma}\label{prop2}
The action of the monodromy matrix element $T_{1,N+1}(z)$ and the zero modes $T_{i+1,i}[0]$ onto off-shell Bethe vectors
$ \mathbb{B}(\bar t)$ are given by the formulas
\begin{equation}\label{ac3}
T_{1,N+1}(z)\mathbb{B}(\bar t) =\lambda_{N+1}(z) \mathbb{B}(\bar w),
\end{equation}
where $\bar w=\{\bar w^1,\ldots,\bar w^N\}$, $\bar w^s = \{z, \bar t^s\}$
and
\begin{equation}\label{ac4}
T_{i+1,i}[0]\mathbb{B}(\bar t)=\sum_{\ell=1}^{r_i}
\sk{\kappa_{i+1}\ \frac{\alpha_i( t^i_\ell)f(\bar t^i_\ell, t^i_\ell)}{f(\bar t^{i+1},t^i_\ell)}
-\kappa_{i}\ \frac{f( t^i_\ell,\bar t^i_\ell)}{f( t^i_\ell,\bar t^{i-1})} }\mathbb{B}(\bar t\setminus \{t^i_\ell\}).
\end{equation}
\end{lemma}

The proof of lemma~\ref{prop2}
can be easily done in the framework of the
current algebra approach for the off-shell
 Bethe vectors in the $\mathfrak{gl}(N+1)$-invariant integrable models (see paper \cite{HLPRS17}).
 If the twisting parameters $\kappa_i=1$,
 $\forall i=1,\ldots,N+1$
 and the Bethe parameters satisfy the  $\mathfrak{gl}(N+1)$-invariant Bethe equations \r{BE}, then
 the on-shell Bethe vectors become $\mathfrak{gl}(N+1)$ highest weight vectors.
Note that the first proof of relations \r{ac3} and \r{ac4}  can be found in \cite{HLPRS17}.
\medskip

Then, the main result of this section is the following
\begin{prop}\label{prop1}
Let $\bar z=\{z_1,\ldots,z_p\}$ be a  set of arbitrary complex variables of  cardinality $\#\bar z=p$.
Then, the multiple action of monodromy matrix elements $T_{i,j}(\bar z)$ onto the off-shell Bethe vector $\BB(\bar t)$ is given by the expression\footnote{For  $i\leq j$, we have
$\prod_{s=j}^{i-1} f(\bar w^s_{\so},\bar w^s_{\sth})=1$ and
$\prod_{s=j}^{i-2} f(\bar w^{s+1}_{\so},\bar w^s_{\sth})=1$, because one of the sets in the arguments of the function $f(u,v)$ is empty.}
\begin{equation}\label{ac2}
\begin{split}
  T_{i,j}(\bar z) \mathbb{B}(\bar t) &=  \lambda_{N+1}(\bar z)
      \sum_{{\rm part}} \mathbb{B}(\bar w_{\st} )
      \frac{\prod_{s=j}^{i-1} f(\bar w^s_{\so},\bar w^s_{\sth})} {\prod_{s=j}^{i-2} f(\bar w^{s+1}_{\so},\bar w^s_{\sth}) }   \\
   &\quad\times    \prod_{s=1}^{i-1} \frac{K(\bar w^s_{\so}| \bar w^{s-1}_{\so}) f(\bar w^s_{\so},\bar w^s_{\st})}
   {f(\bar w^s_{\so}, \bar w^{s-1}_{\so}) f(\bar w^s_{\so},\bar w^{s-1}_{\st})}
		\prod_{s=j}^{N}      \frac{\alpha_{s}(\bar w^{s}_{\sth}) K(\bar w^{s+1}_{\sth}|\bar w^{s}_{\sth})
		 f(\bar w^s_{\st},\bar w^s_{\sth})   }
		{f(\bar w^{s+1}_{\sth}, \bar w^{s}_{\sth})f(\bar w^{s+1}_{\st}, \bar w^{s}_{\sth})}.
		      \end{split}
\end{equation}
 The sum in \r{ac2} runs over partitions obeying the $(i,j)$-condition w.r.t. $\bar z$, and
 $T_{i,j}(\bar z)$ is defined as in \r{shpr}.
\end{prop}

 \textsl{Remark.} It follows from \r{IzD} that the Izergin determinant has poles if $x_i=y_j$, $i,j=1,\dots,p$. Since the intersection of the
sets $\bar w^s$ for different $s$ may be not empty, we can have singularities in the determinants $K(\bar w^s_{\so}| \bar w^{s-1}_{\so})$
or $K(\bar w^{s+1}_{\sth}| \bar w^{s}_{\sth})$. It is easy to see, however, that all these determinants are divided by products of $f$-functions
that compensate these singularities. Strictly speaking, such expressions should be understood as limits, but for brevity we omit the limit symbol.

The proof of proposition~\ref{prop1} is given in appendix~\ref{ApB}. It uses an induction over
$p$ and summation formulas for the Izergin determinant.

\begin{cor} \label{cor1}
The multiple action of monodromy matrix elements $T_{j,i}(\bar z)$ onto the dual off-shell Bethe vectors $\mathbb{C}(\bar t)$ is given by the expression
\begin{equation}\label{dac2}
\begin{split}
\mathbb{C}(\bar t)T_{j,i}(\bar z) &=  \lambda_{N+1}(\bar z)
\sum_{{\rm part}} \mathbb{C}(\bar w_{\st} )
\frac{\prod_{s=j}^{i-1} f(\bar w^s_{\so},\bar w^s_{\sth})} {\prod_{s=j}^{i-2} f(\bar w^{s+1}_{\so},\bar w^s_{\sth}) }   \\
&\quad\times    \prod_{s=1}^{i-1} \frac{K(\bar w^s_{\so}| \bar w^{s-1}_{\so}) f(\bar w^s_{\so},\bar w^s_{\st})}
{f(\bar w^s_{\so}, \bar w^{s-1}_{\so}) f(\bar w^s_{\so},\bar w^{s-1}_{\st})}
\prod_{s=j}^{N}      \frac{\alpha_{s}(\bar w^{s}_{\sth}) K(\bar w^{s+1}_{\sth}|\bar w^{s}_{\sth})
	f(\bar w^s_{\st},\bar w^s_{\sth})   }
{f(\bar w^{s+1}_{\sth}, \bar w^{s}_{\sth})f(\bar w^{s+1}_{\st}, \bar w^{s}_{\sth})}.
\end{split}
\end{equation}
The sum over partitions is taken in the same way as in \r{ac2}.
\end{cor}

The proof of corollary~\ref{cor1} follows directly from the applying antimorphism \r{sp1} to the action formula \r{ac2} .

\section{Scalar products}\label{spsect}

Scalar products of the off-shell Bethe vectors can be written as \cite{HLPRS17a}
\begin{equation}\label{sp6}
S(\bar x|\bar t)=\CC(\bar x)\BB(\bar t)=\sum_{{\rm part}}
Z(\bar x_{\so}|\bar t_{\so})Z(\bar t_{\st}|\bar x_{\st})
\frac{\prod_{j=1}^N\alpha_j(\bar x^j_{\so})\alpha_j(\bar t^j_{\st}) f(\bar x^j_{\st},\bar x^j_{\so}) f(\bar t^j_{\so},\bar t^j_{\st}) }
{\prod_{j=1}^{N-1} f(\bar x^{j+1}_{\st},\bar x^j_{\so}) f(\bar t^{j+1}_{\so},\bar t^j_{\st})},
\end{equation}
where the sum runs over partitions of the sets $\{\bar t^j_{\so},\bar t^j_{\st}\}\vdash \bar t^j$,
 $\{\bar x^j_{\so},\bar x^j_{\st}\}\vdash \bar x^j$ such that $\#\bar t^j_{\so}=\#\bar x^j_{\so}$ for
 all $j=1,\ldots,N$. Function $Z(\bar x|\bar t)$ is known as the highest coefficient of the scalar
 product. It is determined only by the structure of the $R$-matrix entering the commutation relation
 of monodromy matrices. It is normalized in such a way that $Z(\varnothing|\varnothing)=1$.
 We will refer to this formula as the sum formula.
 Let us remark that \cite{Res86} presents the sum formula for the $\mathfrak{gl}(3)$ algebra, the first
 formula of this type was obtained by Korepin
 in \cite{Kor82} for the $\mathfrak{gl}(2)$ algebra.

Note the antimorphism \r{sp2} implies invariance of the scalar product with respect to the replacement $\bar x\leftrightarrow\bar t$:
\begin{equation}\label{psi-sp6}
S(\bar x|\bar t)=\Psi\bigl(S(\bar x|\bar t)\bigr)=\Psi\bigl(\CC(\bar x)\BB(\bar t)\bigr)=\CC(\bar t)\BB(\bar x)=S(\bar t|\bar x).
\end{equation}
This invariance is also easy to see directly from \r{sp6}.

Due to the action formula \r{ac2}  we can obtain recursions for the highest coefficients
with respect to the rank of the algebra.
To describe these recursions we equip the
highest coefficients with an additional subscript: $Z_m(\bar x|\bar t)$. This subscript $m$ means that the
corresponding highest coefficient is related to the scalar product of the Bethe vectors in the $\mathfrak{gl}(m+1)$-invariant quantum integrable model.
Similar subscript  with the same meaning will be used to denote Bethe vectors $\BB_m(\bar t)$ and $\CC_m(\bar x)$ and their scalar products $S_m(\bar x|\bar t)$.

\begin{prop}\label{Prop-HC}
The highest coefficient of the scalar product \r{sp6} satisfies the following recursion:
\begin{equation}\label{RecHC1-par2}
Z_N(\bar x|\bar t)=
\frac{f(\bar t^1,\bar x^1)}{f(\bar t^2,\bar t^1)}
\sum_{{\rm part}} Z_{N-1}(\bar w^2_{\st},\dots,\bar w^N_{\st}|\bar t^2,\dots,\bar t^N)
\prod_{s=1}^{N}      \frac{ K(\bar w^{s+1}_{\sth}|\bar w^{s}_{\sth})
		 f(\bar w^s_{\st},\bar w^s_{\sth})   }
		{f(\bar w^{s+1}, \bar w^{s}_{\sth})}.
\end{equation}
Here
$\bw^s=\{\bar t^1,\bar x^s\}$ for $s=2,\dots,N$. The sum is taken over partitions of $\{\bw^s_{\st},\bw^s_{\sth}\}\vdash\bw^s$ for $s=2,\dots,N$ such that
$\#\bw^s_{\sth}=r_1$. By definition $\bw^1=\bw^1_{\sth}=\bar x^1$ and  $\bw^{N+1}=\bw^{N+1}_{\sth}=\bar t^1$.

Another recursion for the highest coefficient reads
\begin{equation}\label{RecHC2-par3}
Z_N(\bar x|\bar t)=
\frac{f(\bar t^N,\bar x^N)}{f(\bar x^N,\bar x^{N-1})}
\sum_{{\rm part}} Z_{N-1}(\bar x^1,\dots,\bar x^{N-1}|\bar w^1_{\st},\dots,\bar w^{N-1}_{\st})
    \prod_{s=1}^{N} \frac{K(\bar w^s_{\so}| \bar w^{s-1}_{\so}) f(\bar w^s_{\so},\bar w^s_{\st})}
   {f(\bar w^s_{\so}, \bar w^{s-1})}.
\end{equation}
Here $\bar w^s=\{\bar x^N,\bar t^s\}$ for $s=1,\dots,N-1$.
The sum is taken over partitions $\{\bar w^s_{\so},\bar w^s_{\st}\}\vdash \bar w^s$ for $s=1,\dots,N-1$ such that
$\#\bw^s_{\so}=r_N$. By definition $\bar w^0=\bar w^0_{\so}=\bar x^N$ and $\bar w^N=\bar w^N_{\so}=\bar t^N$.
\end{prop}

\proof To prove proposition~\ref{Prop-HC} we use a generalized model.
The notion of the generalized model was introduced in \cite{Kor82} for $\mathfrak{gl}(2)$ based models (see also
\cite{Kor82a,IzeK83,BogIK93L,Res86,HLPRS17b}). This model also can be considered in the case of the quantum integrable models with $\mathfrak{gl}(N+1)$-invariant $R$-matrix.
In fact, the generalized model is a class of models. Each representative of this class has a monodromy matrix
satisfying the $RTT$-relation \eqref{rtt} with the $R$-matrix \eqref{Rmat}, and possesses
reference states $\rvac$ and $\lvac$ with the properties \eqref{tii}, \eqref{sp3}. A representative of the generalized model can be characterized by
a set of the functional parameters $\lambda_i(u)$ \eqref{tii}. Different representatives are distinguished by different
sets of the ratios $\alpha_i(u)$ \eqref{ratios}.

We first prove recursion \r{RecHC1-par2}. Since the highest coefficient is completely determined by the $R$-matrix, it does not depend on the specific choice of the representative of the generalized model. In other words, it does not depend on the free functional parameters $\lambda_i(u)$. Therefore, it enough to prove \r{RecHC1-par2} for some specific choice of $\lambda_i(u)$. We choose them in such a way that
 \be{lambda}
 \lambda_s(t^s_j)=0, \qquad s=1,\dots,N,\quad j=1,\dots,r_s,
 \ee
for given set $\bar t$. This implies
 \be{alpha}
 \alpha_s(t^s_j)=0, \qquad s=1,\dots,N,\quad j=1,\dots,r_s.
 \ee
Then all the subsets $\bar t^j_{\st}$ in \eqref{sp6} are empty. Hence, $\bar t^j_{\so}=\bar t^j$. Since $\#\bar t^j_{\so}=\#\bar x^j_{\so}$, we conclude that
$\bar x^j_{\so}=\bar x^j$. The scalar product then reduces to the highest coefficient
\be{SP-HC}
S(\bar x|\bar t)=Z_N(\bar x|\bar t)\prod_{j=1}^N\alpha_j(\bar x^j).
\ee

Consider the following expectation value:
\be{SPrep1}
Q_N(\bar x|\bar t)=\mathbb{C}_{N}(\varnothing,\bar t^2,\dots,\bar t^N)
\frac{ T_{2,1}(\bar t^1)\BB_N(\bar x)}{\lambda_2(\bar t^1)f(\bar t^2,\bar t^1)}.
\ee
Here $\#\bar t^1=\#\bar x^1=r_1$. Note that  the dual vector $\mathbb{C}_{N}(\varnothing,\bar t^2,\dots,\bar t^N)$ does not depend on the first set of the Bethe parameters. Thus, this vector actually corresponds to a model with $\mathfrak{gl}(N)$-invariant $R$-matrix: $\mathbb{C}_{N}(\varnothing,\bar t^2,\dots,\bar t^N)=\mathbb{C}_{N-1}(\bar t^2,\dots,\bar t^N)$.

The expectation value $Q_N(\bar x|\bar t)$ can be computed in two different ways: either applying the product $T_{2,1}(\bar t^1)$ to the left vector via \eqref{dac2}, or applying $T_{2,1}(\bar t^1)$ to the right vector via \eqref{ac2}. Using the first way we obtain
\begin{equation}\label{act-T12}
Q_N(\bar x|\bar t)=  \frac{\lambda_{N+1}(\bar t^1)}{\lambda_{2}(\bar t^1)f(\bar t^2,\bar t^1)}
      \sum_{{\rm part}} \mathbb{C}_N(\bar w_{\st} )\BB_N(\bar x)
		\prod_{s=2}^{N}      \frac{\alpha_{s}(\bar w^{s}_{\sth}) K(\bar w^{s+1}_{\sth}|\bar w^{s}_{\sth})
		 f(\bar w^s_{\st},\bar w^s_{\sth})   }
		{f(\bar w^{s+1}_{\sth}, \bar w^{s}_{\sth})f(\bar w^{s+1}_{\st}, \bar w^{s}_{\sth})},
    \end{equation}
where $\bar w^1=\bar t^1$, and $\bar w^{s}=\{\bar t^1,\bar t^{s}\}$ for $s=2,\dots,N$. We also impose
the following conditions: $\#\bar w^{s}_{\sth}=r_1$, (for $s=2,\dots,N$), $\bar w^{1}_{\st} =\bar t^1$, $\bar w^{N+1}_{\sth} =\bar t^1$, $\bar w^{N+1}_{\st} =  \varnothing$.

We see that due to \eqref{alpha} $\bar w^{s}_{\sth}=\bar t^1$ for all $s=2,\dots,N$, otherwise we obtain vanishing contributions.
Hence, $\bar w^{s}_{\st}=\bar t^s$ for all $s=2,\dots,N$. The sum over partitions
disappears and we arrive at
\begin{equation}\label{act-T12res}
Q_N(\bar x|\bar t) =  \frac{\lambda_{N+1}(\bar t^1)}{\lambda_{2}(\bar t^1)}
\mathbb{C}_N(\bar t )\BB_N(\bar x)\prod_{s=2}^{N} \alpha_{s}(\bar t^{1})=\mathbb{C}_N(\bar t )\BB_N(\bar x).
    \end{equation}
{\sl Remark.} To obtain \r{act-T12res} from \r{act-T12}, we used the following property of the Izergin determinant
\begin{equation}\label{Izpr}
\lim_{\bar x\to\bar y}\frac{K(\bar x|\bar y)}{f(\bar x,\bar y)}=1.
\end{equation}

Thus, the expectation value $Q_N(\bar x|\bar t)$ is equal to the scalar product $S_N(\bar x|\bar t)$.  Using  \r{SP-HC} we obtain
\begin{equation}\label{mact-T12}
Q_N(\bar x|\bar t) =Z_N(\bar x|\bar t)\prod_{j=1}^N\alpha_j(\bar x^j).
    \end{equation}

Acting with $T_{2,1}(\bar t^1)$ on $\BB_N(\bar x)$ via \eqref{ac2} we obtain
\be{ActT21}
T_{2,1}(\bar t^1)\BB_N(\bar x)=\lambda_{N+1}(\bar t^1)\sum_{{\rm part}} G_{2,1}({\rm part}) \mathbb{B}_{N}(\bar w_{\st} ),
\ee
where $G_{2,1}({\rm part})$ is a numerical coefficient in \eqref{ac2} for $i=2$ and $j=1$. Due to the condition $\#\bar t^1=\#\bar x^1=r_1$ the subset
$\bar w^1_{\st}$ in the resulting vector $\mathbb{B}_{N}(\bar w_{\st} )$ is empty. Therefore, this vector corresponds to $\mathfrak{gl}(N)$-invariant models:
$\mathbb{B}_{N}(\bar w_{\st} )=\mathbb{B}_{N-1}(\bar w^2_{\st},\dots,\bar w^N_{\st} )$. Then we obtain
\be{SPrep1-1}
S_N(\bar x|\bar t)=
\frac{\lambda_{N+1}(\bar t^1)}{\lambda_2(\bar t^1)f(\bar t^2,\bar t^1)}
\sum_{{\rm part}} G_{2,1}({\rm part}) S_{N-1}(\bar w^2_{\st},\dots,\bar w^N_{\st}|\bar t^2,\dots,\bar t^N).
\ee
Taking into account \eqref{SP-HC} we arrive at
\be{RecHC1-gen}
Z_N(\bar x|\bar t)\prod_{j=1}^N\alpha_j(\bar x^j)=
\frac{\lambda_{N+1}(\bar t^1)}{\lambda_2(\bar t^1)f(\bar t^2,\bar t^1)}
\sum_{{\rm part}} G_{2,1}({\rm part}) Z_{N-1}(\bar w^2_{\st},\dots,\bar w^N_{\st}|\bar t^2,\dots,\bar t^N)\prod_{j=2}^N\alpha_j(\bar w^j_{\st}).
\ee
It remains to use the explicit form of $G_{2,1}({\rm part})$. Setting $i=2$ and $j=1$ in \eqref{ac2} we find
\begin{multline}\label{RecHC1-par1}
Z_N(\bar x|\bar t)\prod_{j=1}^N\alpha_j(\bar x^j)=
\frac{\lambda_{N+1}(\bar t^1)}{\lambda_2(\bar t^1)f(\bar t^2,\bar t^1)}
\sum_{{\rm part}} Z_{N-1}(\bar w^2_{\st},\dots,\bar w^N_{\st}|\bar t^2,\dots,\bar t^N)\prod_{j=2}^N\alpha_j(\bar w^j_{\st})\\
\times    \frac{K(\bar w^1_{\so}| \bar t^1) f(\bar w^1_{\so},\bar w^1_{\sth})}
   {f(\bar w^1_{\so}, \bar t^1) }
		\prod_{s=1}^{N}      \frac{\alpha_{s}(\bar w^{s}_{\sth}) K(\bar w^{s+1}_{\sth}|\bar w^{s}_{\sth})
		 f(\bar w^s_{\st},\bar w^s_{\sth})   }
		{f(\bar w^{s+1}_{\sth}, \bar w^{s}_{\sth})f(\bar w^{s+1}_{\st}, \bar w^{s}_{\sth})}.
\end{multline}
Here $\bw^s=\{\bar t^1,\bar x^s\}$ for $s=2,\dots,N$. The sum is taken over partitions $\{\bw^s_{\st},\bw^s_{\sth}\}\vdash\bw^s$ for $s=2,\dots,N$ such that
$\#\bw^s_{\sth}=r_1$. The subsets $\bw^1_{\so}$, $\bw^1_{\st}$, and $\bw^1_{\sth}$ actually are fixed by the condition \eqref{alpha}:
$\bw^1_{\so}=\bar t^1$, $\bw^1_{\sth}=\bar x^1$, and $\bw^{1}_{\st}=\varnothing$. Finally, by definition $\bw^{N+1}_{\sth}=\bar t^1$ and
$\bw^{N+1}_{\st}=\varnothing$.

First of all, it is easy to see that
\be{prod-al}
\prod_{j=2}^N\alpha_j(\bar w^j_{\st})\prod_{s=1}^{N} \alpha_{s}(\bar w^{s}_{\sth})=
\alpha_1(\bar x^1)\prod_{j=2}^N\alpha_j(\bar w^j)=\frac{\lambda_2(\bar t^1)}{\lambda_{N+1}(\bar t^1)}\prod_{j=1}^N\alpha_j(\bar x^j).
\ee
 Using again the property \r{Izpr}, we also have
\be{using1}
\frac{K(\bar w^1_{\so}| \bar t^1)}{f(\bar w^1_{\so}, \bar t^1) }=
\frac{K(\bar t^1| \bar t^1)}{ f(\bar t^1, \bar t^1)}=1.
\ee
Substituting \r{prod-al} and \r{using1} into \r{RecHC1-par1} we arrive at \r{RecHC1-par2}.
Thus, the recursion \r{RecHC1-par2} is proved.

The second recursion \eqref{RecHC2-par3} follows from \r{RecHC1-par2} due to an isomorphism
$\varphi:\ Y(\mathfrak{gl}(N+1))\ \to\ Y(\mathfrak{gl}(N+1))\Big|_{c\to-c}$ between Yangians with reflected parameters $c\to -c$ \cite{PakRS17,HLPRS17a}.  On the elements
of the monodromy matrix, it is given explicitly by
\begin{equation}\label{morphism}
\varphi\bigl(T_{i,j}(u)\bigr)= \hat T_{N+2-j,N+2-i}(u)\,,\qquad i,j=1,\dots,N+1.
\end{equation}
Here $T_{i,j}(u)\in Y(\mathfrak{gl}(N+1))$ and $\hat T_{i,j}(u)\in Y(\mathfrak{gl}(N+1))\Big|_{c\to-c}$.

However, it is easier to directly derive \eqref{RecHC2-par3} by renormalizing the initial Bethe vectors as follows:
\be{newBV}
\begin{aligned}
&\hBB_N(\bar t)=\BB_N(\bar t)\prod_{s=1}^N\beta_s(\bar t^s),\\
&\hCC_N(\bar t)=\CC_N(\bar t)\prod_{s=1}^N\beta_s(\bar t^s),
\end{aligned}
\qquad\qquad \beta_s(z)=\frac1{\alpha_s(z)}=\frac{\lambda_{s+1}(z)}{\lambda_s(z)}.
\ee
Then it is easy to see that  scalar product of these new Bethe vectors takes the form
\begin{equation}\label{sp7}
\hat S_N(\bar x|\bar t)=\hCC_N(\bar x)\hBB_N(\bar t)=\sum_{{\rm part}}
Z_N(\bar x_{\so}|\bar t_{\so})Z_N(\bar t_{\st}|\bar x_{\st})
\frac{\prod_{j=1}^N\beta_j(\bar x^j_{\st})\beta_j(\bar t^j_{\so}) f(\bar x^j_{\st},\bar x^j_{\so}) f(\bar t^j_{\so},\bar t^j_{\st}) }
{\prod_{j=1}^{N-1} f(\bar x^{j+1}_{\st},\bar x^j_{\so}) f(\bar t^{j+1}_{\so},\bar t^j_{\st})}.
\end{equation}
The sum is taken over partitions as in \eqref{sp6}.

The action formulas \eqref{ac2} and \eqref{dac2} also change. They respectively turn into
\begin{equation}\label{act-det2}
\begin{split}
 T_{i,j}(\bar z) \hBB_N(\bar t) &=  \lambda_{1}(\bar z)
      \sum_{{\rm part}} \hBB_N(\bar w_{\st} )
      \frac{\prod_{s=j}^{i-1} f(\bar w^s_{\so},\bar w^s_{\sth})} {\prod_{s=j}^{i-2} f(\bar w^{s+1}_{\so},\bar w^s_{\sth}) }   \\
   &\quad\times    \prod_{s=1}^{i-1} \frac{\beta_{s}(\bar w^{s}_{\so})K(\bar w^s_{\so}| \bar w^{s-1}_{\so}) f(\bar w^s_{\so},\bar w^s_{\st})}
   {f(\bar w^s_{\so}, \bar w^{s-1}_{\so}) f(\bar w^s_{\so},\bar w^{s-1}_{\st})}
		\prod_{s=j}^{N}      \frac{ K(\bar w^{s+1}_{\sth}|\bar w^{s}_{\sth})
		 f(\bar w^s_{\st},\bar w^s_{\sth})   }
		{f(\bar w^{s+1}_{\sth}, \bar w^{s}_{\sth})f(\bar w^{s+1}_{\st}, \bar w^{s}_{\sth})},
		      \end{split}
    \end{equation}
and
\begin{equation}\label{dact-det2}
\begin{split}
 \hCC_N(\bar t)T_{j,i}(\bar z) &=  \lambda_{1}(\bar z)
      \sum_{{\rm part}} \hCC_N(\bar w_{\st} )
      \frac{\prod_{s=j}^{i-1} f(\bar w^s_{\so},\bar w^s_{\sth})} {\prod_{s=j}^{i-2} f(\bar w^{s+1}_{\so},\bar w^s_{\sth}) }   \\
   &\quad\times    \prod_{s=1}^{i-1} \frac{\beta_{s}(\bar w^{s}_{\so})K(\bar w^s_{\so}| \bar w^{s-1}_{\so}) f(\bar w^s_{\so},\bar w^s_{\st})}
   {f(\bar w^s_{\so}, \bar w^{s-1}_{\so}) f(\bar w^s_{\so},\bar w^{s-1}_{\st})}
		\prod_{s=j}^{N}      \frac{ K(\bar w^{s+1}_{\sth}|\bar w^{s}_{\sth})
		 f(\bar w^s_{\st},\bar w^s_{\sth})   }
		{f(\bar w^{s+1}_{\sth}, \bar w^{s}_{\sth})f(\bar w^{s+1}_{\st}, \bar w^{s}_{\sth})}.
		      \end{split}
    \end{equation}
The partitions are the same as in \eqref{ac2} and \eqref{dac2}.

Now to derive recursion \eqref{RecHC2-par3} we take a new representative of the generalized model, for which
 \be{lambda-n}
 \lambda_{s+1}(t^s_j)=\beta_{s}(t^s_j)=0, \qquad s=1,\dots,N,\quad j=1,\dots,r_s.
 \ee
Thus,
\begin{equation}\label{sp-HCn}
\hat S_N(\bar x|\bar t)= Z_N(\bar t|\bar x)\prod_{s=1}^N\beta_s(\bar x^s).
\ee

In complete analogy with the derivation of recursion \eqref{RecHC1-par2} we consider the following expectation value
\be{SPrep2}
\hat Q_N(\bar x|\bar t)=\hCC_{N}(\bar t^1,\dots,\bar t^{N-1},\varnothing)
\frac{ T_{N+1,N}(\bar t^N)\BB_N(\bar x)}{\lambda_N(\bar t^N)f(\bar t^{N},\bar t^{N-1})},
\ee
where $\#\bar t^N=\#\bar x^N=r_N$.
After that, we repeat the calculations already done. Acting with the product $T_{N+1,N}(\bar t^N)$ to the left we obtain
$\hat Q_N(\bar x|\bar t)=\hat S_N(\bar x|\bar t)$. The action of the same product to the right gives us the desired recursion
with relabeling $\bar x  \leftrightarrow \bar t$. \qed

In paper \cite{LPRS19}, we found a symmetry of  the highest coefficient with respect to a special reordering and shifts of the Bethe parameters.
To describe this symmetry we introduce a mapping
\begin{equation}\label{map}
\mu(\bar t)\equiv \mu(\{\bar t^1,\bar t^2,\dots,\bar t^{N}\})=\{\bar t^{N},\bar t^{N-1} - c,\dots,\bar t^{1}-(N-1)c\}.
\end{equation}
Then it follows from the results of \cite{LPRS19} that
\begin{equation}\label{ZZ1}
Z_N(\bar x|\bar t)
\prod_{k=1}^{N-1}  f(\bar x^{k+1},  \bar x^{k})f(\bar t^{k+1},  \bar t^{k})
= Z_N\bigl(\mu(\bar x)|\mu(\bar t)\bigr).	
\end{equation}
Equation \r{ZZ1} and recursions \r{RecHC1-par2} and \r{RecHC2-par3} imply two more recursions  for the highest coefficient.

\begin{cor}\label{cor-HC}
The highest coefficient of the scalar product \r{sp6} satisfies two more recursions
\begin{multline}\label{sp10aa}
Z_N(\bar x|\bar t)=\frac{(-1)^{r_1N}f(\bar t^1,\bar x^{1})}{\prod_{k=1}^{N-1}f(\bar t^{k+1},\bar t^{k})}
\sum_{{\rm part}} Z_{N-1}(\bar x^2,\dots, \bar x^N|\bar\eta^{2}_{\st},\dots,\bar\eta^{N}_{\st})\\
\times\prod_{s=1}^N K(\bar\eta^{s+1}_{\so}| \bar\eta^{s}_{\so}) f(\bar\eta^s_{\so},\bar\eta^s_{\st})
   f(\bar\eta^{s+1}_{\st}, \bar\eta^{s}) .\qquad
\end{multline}
Here the sum runs over partitions of the sets
$\{\bar\eta^s_{\so},\bar\eta^s_{\st}\}\vdash \bar\eta^s=\{\bar t^s,\bar x^1-(s-1)c\}$
for $s=2,\ldots,N$ such that $\#\bar\eta^s_{\so}=r_1$ and $\bar\eta^{N+1}_{\so}=\bar x^1-Nc$, $\bar\eta^1_{\so}=\bar t^1$,
$\bar\eta^1_{\st}=\bar\eta^{N+1}_{\st}=\varnothing$.
\begin{multline}\label{sp10ab}
Z_N(\bar x|\bar t)=\frac{(-1)^{r_NN}f(\bar t^N,\bar x^{N})}{\prod_{k=1}^{N-1}f(\bar x^{k+1},\bar x^{k})}
\sum_{{\rm part}} Z_{N-1}(\bar\eta^{1}_{\st},\dots,\bar\eta^{N-1}_{\st}|\bar t^1,\dots, \bar t^{N-1})\\
\times\prod_{s=1}^N K(\bar\eta^{s}_{\sth}| \bar\eta^{s-1}_{\sth}) f(\bar\eta^s_{\st},\bar\eta^s_{\sth})
   f(\bar\eta^{s}, \bar\eta^{s-1}_{\st}) .\qquad
\end{multline}
Here the sum runs over partitions of the sets
$\{\bar\eta^s_{\st},\bar\eta^s_{\sth}\}\vdash \bar\eta^s= \{\bar x^s,\bar t^{N}-(N-s)c\}$
for $s=1,\ldots,N-1$ such that $\#\bar\eta^s_{\sth}=r_N$ and $\bar\eta^{N}_{\sth}=\bar x^N$, $ \bar\eta^0_{\sth}=\bar t^N+Nc$,
$\bar\eta^0_{\st}=\bar\eta^{N}_{\st}=\varnothing$.
\end{cor}
To prove these recursions it is enough to apply \eqref{RecHC2-par3} and \eqref{RecHC1-par2}
to $Z_N\bigl(\mu(\bar x)|\mu(\bar t)\bigr)$.	
\qed

Note that the use of both recursions \r{RecHC2-par3} and \r{sp10aa} played a very important role in deriving determinant
representations for the scalar products in $\mathfrak{gl}(3)$-invariant models \cite{BelPRS12a}.
In the $\mathfrak{gl}(2)$ case, the four recursions coincide and give 
$Z_1(\bar x|\bar t)=K(\bar t|\bar x)$, as expected from \cite{Kor82}.

\section{Results for $\mathfrak{gl}(m|n)$ related models}\label{super}

In this section, we use notation of the paper \cite{HLPRS17} to describe monodromy matrices
as the matrices acting in the auxiliary space  $\CC^{m|n}$. This is a $\ZZ_2$-graded space with a basis ${\rm e}_i$, $i=1,\ldots,m+n$.
We assume that the basis vectors $\{{\rm e}_1,{\rm e}_2,\ldots, {\rm e}_m\}$ are even
while $\{{\rm e}_{m+1},{\rm e}_{m+2},\ldots, {\rm e}_{m+n}\}$ are odd. In this section  $N= m+n -1$.
We introduce the $\ZZ_2$-grading of the indices as\footnote{We use the same notation $\prt{i}$ to describe the parity function and to define modes of the generating series in \r{depen}. However, in this section we will use only zero modes operators marked by the symbol $[0]$, so that the distinction with the parity function will be clear enough since 0 is not an index of \eqref{grade}.}
\begin{equation}\label{grade}
\prt{i}=0\quad\mbox{for}\quad i=1,2,\ldots,m,\quad\mbox{and}\quad \prt{i}=1\quad\mbox{for}\quad i=m+1,m+2,\ldots,m+n.
\end{equation}

Let $\EE_{ij}\in{{\textrm{End}}(\CC^{m|n})}$ be again a matrix with the only nonzero entry
equal to $1$ at the intersection of the $i$-th row and $j$-th column. Monodromy matrix elements are
defined by the same formula \r{mmel} but now the matrices $\EE_{ij}$ have grading
\begin{equation*}
 [\EE_{ij}]=\prt{i}+\prt{j} {\mod 2}.
\end{equation*}
 Their tensor products are also graded according to the rule
\begin{equation*}
(\EE_{ij}\otimes\EE_{kl})\cdot (\EE_{pq}\otimes\EE_{rs})
=(-)^{(\prt{k}+\prt{l})(\prt{p}+\prt{q})} \EE_{ij}\EE_{pq}\otimes \EE_{kl}\EE_{rs}\,,
\end{equation*}
as well as the transposition antimorphism:
\begin{equation}
\Psi\ :\ T_{i,j}(u)\to (-1) ^{[i]([j]+1)}\,T_{j,i}(u),\qquad \Psi(A\cdot B)=(-1) ^{[A][B]}\,\Psi(B)\cdot \Psi(A).
\end{equation}

The commutation relations of the monodromy matrices $T(u)$ are the same as in \r{rtt} with
the same structure of the $R$-matrix as in \r{Rmat}, but with graded
permutation operator  $\mathbf{P}$  acting in the tensor product $\CC^{m|n}\otimes \CC^{m|n}$ as
\begin{equation*}
{\rm P}=\sum_{i,j=1}^{m+n}(-)^{\prt{j}}\ \EE_{ij}\otimes \EE_{ji}.
\end{equation*}
This results in slightly different commutation relations for the monodromy matrix elements \cite{HLPRS17}.

 To describe the action of the graded monodromy matrix elements onto supersymmetric Bethe vectors,
it is convenient to introduce `colored' analogs the rational functions $g(u,v)$, $f(u,v)$, and $h(u,v)$ \r{f}:
\begin{equation}\label{f-sup}
f_{\prt{i}}(u,v)=1+g_{\prt{i}}(u,v)=1+\frac{\cci{i}}{u-v}=\frac{u-v+\cci{i}}{u-v}\,,\qquad
 h_{\prt{i}}(u,v)=\frac{f_{\prt{i}}(u,v)}{g_{\prt{i}}(u,v)},
\end{equation}
where
\begin{equation*}
c_{\prt{i}}=(-)^{\prt{i}}c\,.
\end{equation*}

Second,  for arbitrary sets of  parameters
$\bu$ and $\bv$ we define
\begin{equation}\label{dc9}
\Fli{i}(\bu,\bv)=\frac{f_{\prt{i}}(\bu,\bv)}{h(\bu,\bv)^{\delta_{i,m}}}\quad\mbox{and}
\quad \hFli{i}(\bu,\bv)=\frac{f_{\prt{i+1}}(\bu,\bv)}{h(\bv,\bu)^{\delta_{i,m}}}\,.
\end{equation}
The first  function in \r{dc9} coincides with the function $f_{\prt{i}}(\bu,\bv)$ for $i\not=m$ and with
$g(\bu,\bv)$ for $i=m$, while the second function coincides with
$f_{\prt{i+1}}(\bu,\bv)$ for $i\not=m$ and with
$g(\bv,\bu)$ for $i=m$. Note that
\begin{equation}\label{equ}
\Fli{m}(\bu,\bv)=(-)^{\#\bu\cdot\#\bv} \hFli{m}(\bu,\bv)\,.
\end{equation}
and $\Fli{i}(\bu,\bv) = \hFli{i}(\bu,\bv)$ for $i\not=m$.

Then lemma~\ref{prop2} is replaced by
\begin{lemma}\label{p2su}
The action of the monodromy matrix element $T_{1,N+1}(z)$ and the zero modes $T_{i+1,i}[0]$ onto off-shell Bethe vectors
$ \mathbb{B}(\bar t)$ are given by the formulas
\begin{equation}\label{susy1}
T_{1,N+1}(z)\mathbb{B}(\bar t) =\lambda_{N+1}(z) h(\bar t^m,z) \mathbb{B}(\bar w),
\end{equation}
and
\begin{equation}\label{susy2}
T_{i+1,i}[0]\mathbb{B}(\bar t)=(-1)^{\prt{i+1}}\sum_{\ell=1}^{r_i}
\sk{\kappa_{i+1}\ \frac{\alpha_i(t^i_\ell)\Fli{i}(\bar t^i_\ell, t^i_\ell)}{f_{\prt{i+1}}(\bar t^{i+1},t^i_\ell)}
-\kappa_{i}\ \frac{\hFli{i}( t^i_{\ell},\bar t^i_{\ell})}{f_{\prt{i}}( t^i_{\ell},\bar t^{i-1})} }\mathbb{B}(\bar t\setminus \{t^i_\ell\}).
\end{equation}
\end{lemma}

The supersymmetric Bethe equations
\begin{equation}\label{BEsup}
 \alpha_i(t^i_\ell)=\frac{\hFli{i}(t^i_\ell,\bar t^i_\ell)}{\Fli{i}(\bar t^i_\ell,t^i_\ell)}\
 \frac{f_{\prt{i+1}}(\bar t^{i+1},t^i_\ell)}{f_{\prt{i}}(t^i_\ell,\bar t^{i-1})}
\end{equation}
provide (when all $\kappa_i=1$) the highest weight condition for the Bethe vectors $\BB(\bar t)$ with respect to the
raising operators of the finite dimensional algebra $\mathfrak{gl}(m|n)$ formed by the
zero modes operators. Due to the properties \r{equ} the supersymmetric Bethe equation for
the Bethe parameters $t^m_\ell$ (corresponding to the odd simple root) simplifies to
\begin{equation*}
 \alpha_m(t^m_\ell)=
 \frac{f(t^m_\ell,\bar t^{m+1})}{f(t^m_\ell,\bar t^{m-1})}\,.
\end{equation*}

To describe the multiple action in the supersymmetric case we introduce the following symmetric products:
 \begin{equation}
	\mathbb{T}_{i,j}(\bar z) =
	\begin{cases}
	  T_{i,j}(\bar z), &\text{if  $[i]+[j]=0\ {\rm mod}\ 2$},\\
	  \Delta_h(\bar z)^{-1}T_{i,j}(z_1)\cdots T_{i,j}(z_p), &\text{if $[i]=0$ and $[j]=1$},\\
	   \Delta'_h(\bar z)^{-1}T_{i,j}(z_1)\cdots T_{i,j}(z_p), &\text{if $[i]=1$ and $[j]=0$.}
	\end{cases}
      \end{equation}
According to the commutation relations between monodromy matrix elements in the supersymmetric
case, the product $\mathbb{T}_{i,j}(\bar z)$ is symmetric with respect to any permutation in the set $\bar z$.

Proposition~\ref{prop1} is replaced in the supersymmetric case by
\begin{prop}\label{p1susy}
The action by $\mathbb{T}_{i,j}(\bar z)$
onto supersymmetric off-shell Bethe vector $\BB(\bar t)$ is
\begin{equation}\label{Act}
\begin{split}
	 \mathbb{T}_{i,j}(\bar z) \mathbb{B}(\bar t) &=
	      \lambda_{N+1}(\bar z)h(\bar t^m ,\bar z)
	      \prod_{s=j}^{i-1} \frac{(-1)^{(\prt{s}+\prt{s+1})\frac{n^2-n}{2}}}{h(\bar z,\bar z)^{\delta_{s,m}}}
	      \sum_{{\rm part}}   \mathbb{B}(\bar w_{\st} )
	      \frac{\prod_{s=j}^{i-1}  \hFli{s}(\bar w^s_{\so},\bar w^s_{\sth})}
		   {\prod_{s=j}^{i-2} f_{\prt{s+1}}(\bar w^{s+1}_{\so},\bar w^s_{\sth})}	\\
	       &\quad\times\hat{\mathcal{K}}_i \left(\bar w_{\so} \right)
	       \prod_{s=1}^{i-1}   \frac{ \hFli{s}(\bar w^s_{\so},\bar w^s_{\st})}
	 			    { f_{[s]}(\bar w^s_{\so},\bar w^{s-1}_{\st})  }
	      \; \mathcal{K}_j \left(\bar w_{\sth} \right)
	     \prod_{s=j}^{N}   \frac{\alpha_{s}(\bar w^{s}_{\sth})  \gamma_s(\bar w^s_{\st}, \bar w^s_{\sth}) }
	 			    { f_{[s+1]}(\bar w^{s+1}_{\st},\bar w^{s}_{\sth})}.
      \end{split}
      \end{equation}
Here instead of products of the Izergin determinants, we introduce
\begin{equation}\label{susy3}
      \hat{\mathcal{K}}_i \left(\bar w_{\so} \right)  =
      \begin{cases}
         \prod_{s=1}^{i-1} \frac{K_{[s]}(\bar w^{s}_{\so}|\bar w^{s-1}_{\so})}{f_{[s]}(\bar w^{s}_{\so},\bar w^{s-1}_{\so})}
         &\text{if $i\le m$,}\\
         \prod_{s=1}^{m} \frac{h_{[s]}(\bar w^{s-1}_{\so},\bar w^{s-1}_{\so})}{h_{[s]}(\bar w^{s}_{\so},\bar w^{s-1}_{\so})}
         \prod_{s=m+1}^{i-1} \frac{K_{[s]}(\bar w^{s}_{\so}|\bar w^{s-1}_{\so})}{f_{[s]}(\bar w^{s}_{\so},\bar w^{s-1}_{\so})} &\text{if $i>m$,}
      \end{cases}
    \end{equation}
and
 \begin{equation}\label{susy4}
	\mathcal{K}_j \left(\bar w_{\sth} \right) =
	\begin{cases}
	  \prod_{s=j}^{m-1} \frac{K_{[s+1]}(\bar w^{s+1}_{\sth}|\bar w^{s}_{\sth})}{f_{[s+1]}(\bar w^{s+1}_{\sth},\bar w^{s}_{\sth})}
 \prod_{s=m}^{N} \frac{h_{[s+1]}(\bar w^{s+1}_{\sth},\bar w^{s+1}_{\sth})}{h_{[s+1]}(\bar w^{s+1}_{\sth},\bar w^{s}_{\sth})}
 &\text{if $j\le m,$}\\[1.2ex]
	  \prod_{s=j}^{N} \frac{K_{[s+1]}(\bar w^{s+1}_{\sth}|\bar w^{s}_{\sth})}{f_{[s+1]}(\bar w^{s+1}_{\sth},\bar w^{s}_{\sth})}
	  &\text{if $j>m.$}
	\end{cases}
     \end{equation}
In \r{susy3} and \r{susy4} the symbol $K_{\prt{s}}(\bar x|\bar y)$ means the Izergin determinant given
by the expression \r{IzD}, where the functions $g(x,y)$ and $h(x,y)$ are replaced by their graded
analogs $g_{\prt{s}}(x,y)$ and $h_{\prt{s}}(x,y)$ given by \r{f-sup}.

Summations  in  $\r{Act}$  are
over the partitions of the
sets $\{\bar w^s_{\so},\bar w^s_{\st},\bar w^s_{\sth}\}\vdash
\bar w^s=\{\bar z,\bar t^s\}$ that obey the $(i,j)$-condition w.r.t. $\bar z$.
\end{prop}

The proof of proposition~\ref{p1susy} is carried out by the same method  as the one given in the appendices~\ref{ApA} and~\ref{ApB}.  If one specifies $m=0$ or $n=0$, proposition \ref{p1susy} reduces to proposition \ref{prop1}.

These action formulas can be used
to find a recurrence relations for the highest coefficient of the scalar product of the supersymmetric
off-shell Bethe vectors.
In the supersymmetric case the scalar product takes the following form
\begin{equation}\label{al-dep}
S(\bx|\bar t)= \sum_{\text{part}}
Z^{m|n}(\bx_{\so}|\bar t_{\so}) \;\; Z^{m|n}(\bar t_{\st}|\bx_{\st})
\;  \frac{\prod_ {k=1}^{N} \alpha_{k}(\bx^k_{\so}) \alpha_{k}(\bar t^k_{\st}) \gamma_{k}(\bx^k_{\st},\bx^k_{\so}) \gamma_{k}(\bar t^k_{\so},\bar t^k_{\st})}
{\prod_{j=1}^{N-1} f_{[j+1]}(\bar {x}^{j+1}_{\st},\bar {x}^j_{\so})
	f_{[j+1]}(\bar t^{j+1}_{\so},\bar t^j_{\st})}.
\end{equation}
The highest
coefficient\footnote{{Let us note the
difference of notation between $Z_N$ (bosonic case) and  $Z^{m|n}$ (supersymmetric case):
indeed, we have $Z^{m|0}=Z_{m-1}$.}}
$Z^{m|n}(\bar x | \bar t) $ satisfies the following
recursion.

\begin{prop}\label{Prop-HC-susy}
The highest coefficient of the scalar product \r{al-dep} satisfies the recursions
\begin{multline}\label{rec-super1}
Z^{m|n}(\bar x | \bar t)  =  \frac{ \hat\gamma_N(\bar t^N,\bar x^N)  \; h(\bar t^m ,\bar x^{N}) }{ f_{[N]}(\bar x^{N},\bar x^{N-1}) \; h(\bar x^{N},\bar x^{N})^{\delta_{m,N}}}
\\
\times\sum_{\text{part}}
Z^{m|n-1}(\bar x^1,.., \bar x^{N-1} | \bar w_{\st}^1,..,\bar w_{\st}^{N-1})
\; \hat{\mathcal{K}}_{N+1} \left(\bar w_{\so} \right)	
 \prod_{s=1}^{N}
\frac{    \hat\gamma_s(\bar w^s_{\so},\bar w^s_{\st}) }
{  f_{[s]}(\bar w^s_{\so},\bar w^{s-1}_{\st})     },\qquad
\end{multline}
and
\begin{multline}\label{rec-super2}
	Z^{m|n}(\bar x | \bar t)  =  \frac{\hat\gamma_1(\bar t^1,\bar x^1)  \;
		{  h(\bar x^{m},\bar t^1)}}{{ f_{[2]}(\bar t^{2}, \bar t^1)}\, h(\bar t^{1},\bar t^{1})^{\delta_{m,1}}}\;
	\\
	\times\sum_{\text{part}}
	Z^{m-1|n}(\bar w^2_{\st},.., \bar w_{\st}^N \big| \bar t^2 ,.., \bar t^N)
	\; {\mathcal{K}}_{1} \left(\bar w_{\sth} \right)
	\prod_{s=1}^{N}\frac{
		\gamma_s(\bar w^s_{\st},\bar w^s_{\sth}) }{  f_{[s+1]}(\bar w^{s+1}_{\st},\bar w^{s}_{\sth})}.
	\qquad
\end{multline}

The sums run over partitions as in proposition~\ref{Prop-HC}.
The formula \eqref{rec-super1} works only for $n \ge 1$, while recursion \eqref{rec-super2} is valid only when $m\geq1$.
\end{prop}

 Equations \eqref{rec-super1} and \eqref{rec-super2} are related by the symmetry \cite{PakRS17,HLPRS17a}
\begin{equation}\label{ZZmn-phi}
Z^{m|n}(\bar x|\bar t) = (-1)^{r_m} Z^{n|m}\bigl(\olt|\olx\bigr) ,
\end{equation}
with $\olx=(\bar x^N,...,\bar x^2,\bar x^1)$. Relation \eqref{ZZmn-phi}
 comes from the action of the morphism
\begin{equation}\label{phi-sup}
\varphi\ :\ \left\{
\begin{array}{lcl}
Y(m|n)\ &\mapsto & Y(n|m),\\
T_{ij}(u) &\mapsto& (-1)^{[i][j]+[j]}\,\widetilde T_{\bar\jmath \bar\imath}(u),  \quad \bar k = N+2-k, \\
\lambda_j(u)&\mapsto& \lambda_{\bar\jmath}(u), \\
\alpha_j(u)&\mapsto& \alpha_{\bar\jmath}(u)^{-1},
\end{array}\right.
\end{equation}
on the scalar product \eqref{al-dep}.
Let us remark that in connecting $Z^{m|n}$ to $Z^{n|m}$, one gets functions such as $f_{[s]}(x,y)$, which are different if they are associated
to $Y(\mathfrak{gl}(m|n))$ or to $Y(\mathfrak{gl}(n|m))$, since the grading $[.]$ depends on which Yangian one considers. For instance, one has
$$f^{n|m}_{[s+1]}(x,y)=f^{m|n}_{[N+1-s]}(y,x),$$
with obvious notation. In the same way, we have
\begin{equation}
\hat\gamma^{n|m}_s(\bar u, \bar v) = \gamma^{m|n}_{N+1-s}(\bar v,\bar u)
\qquad \text{and}\qquad \hat{\mathcal{K}}^{n|m}_{N+1} (\bar w ) = {\mathcal{K}}^{m|n}_{1} (\bar \eta )\,,\ \text{with }
\bar \eta^s = \bar w^{N+1-s}.
\end{equation}
One has to pay attention to these differences when using the morphism \eqref{phi-sup} on \eqref{rec-super1}.  Again, if one sets $m=0$ or $n=0$, the proposition \ref{Prop-HC-susy} reduces to proposition \ref{Prop-HC}.

\begin{cor}\label{cor-HC-susy}
The highest coefficient of the scalar product \r{al-dep} also satisfies the recursion
\begin{multline}\label{rec-super3}
Z^{m|n}(\bar x | \bar t)  =  (-1)^{N\,r_1}\;
	\frac{\hat\gamma_1(\bar t^1,\bar x^1)  \;	h(\bar x^{1}-(m-1)c,\bar t^m)} { h(\bar x^{1},\bar x^{1})^{\delta_{m,1}}}\;
	\prod_{k=1}^{N-1} f_{[k+1]}(\bar t^{k+1}, \bar t^k)^{-1}
	\\
	\times\sum_{\text{part}}
	Z^{m-1|n}(\bar x^2,.., \bar x^N | \bar \eta_{\st}^2,..,\bar \eta_{\st}^N)
	\; {\mathcal{K}}_{1} \left(\bar \eta_{\so} \right)	
	\prod_{s=1}^{N}
	 \gamma_s(\bar \eta^s_{\so},\bar \eta^s_{\st})   f_{[s+1]}(\bar \eta^{s+1}_{\so},\bar \eta^{s}_{\so}) f_{[s+1]}(\bar \eta^{s+1}_{\st},\bar \eta^{s} ),
	\end{multline}
where $\bar \eta^1 = \bar t^1$,  $\bar \eta^s = \{ \bar t^s, \bar x^1 - (s - 1) c \}$ for $s=2,\ldots,m$,
$\bar \eta^s = \{ \bar t^s, \bar x^1 - (2 m - s - 1) c \}$ for $s=m+1,\ldots,N$, and 
$\bar \eta^{N+1} =  \bar x^1 - (m-n-1) c $. The sum runs over partitions $\{\bar\eta^s_{\so},\bar\eta^s_{\st}\}\vdash \bar\eta^s$ with $\#\bar\eta^s_{\so}=r_1$ and $\bar\eta^1_{\st}=\bar\eta^{N+1}_{\st}=\varnothing$. The formula \eqref{rec-super3} works only for $m \ge 1$.

It obeys also the recursion
\begin{multline}\label{rec-super4}
Z^{m|n}(\bar x | \bar t)  =  (-1)^{N\,r_N}\; \frac{\hat\gamma_N(\bar t^N,\bar x^N)  \;
	h(\bar t^{N}-(n-1)c,\bar x^m)}{ h(\bar t^{N},\bar t^{N})^{\delta_{n,1}}}\;
\prod_{k=2}^{N} f_{[k]}(\bar x^{k}, \bar x^{k-1})^{-1}
\\
\qquad\times\sum_{\text{part}}
Z^{m|n-1}( \bar \eta^1_{\st},..,\bar \eta^{N-1}_{\st} | \bar t^1 ,.., \bar t^{N-1})
\; \hat{\mathcal{K}}_{N+1} \left(\bar \eta_{\sth} \right)	
\prod_{s=1}^{N}
\hat\gamma_s(\bar \eta^s_{\st},\bar \eta^s_{\sth})   f_{[s]}(\bar \eta^{s}_{\sth},\bar \eta^{s-1})
f_{[s]}(\bar\eta^{s}_{\st}, \bar \eta^{s-1}_{\st}),
\end{multline}
where $\bar \eta^0 = \bar t^N-(n-m-1)c$,  $\bar \eta^s = \{ \bar x^s, \bar t^N - (s+n-m - 1) c \}$ for $s=1,\ldots,m$,
$\bar \eta^s = \{ \bar x^s, \bar t^N - ( m+n - 1- s ) c \}$ for $s=m+1,\ldots,N-1$, and 
$\bar \eta^{N} =  \bar x^N $.
The sums run over partitions $\{\bar\eta^s_{\st},\bar\eta^s_{\sth}\}\vdash \bar\eta^s$ with
$\#\bar\eta^s_{\sth}=r_N$ and $\bar\eta^0_{\st}=\bar\eta^{N}_{\st}=\varnothing$. 
This formula  is valid only when $n \ge 1$.
\end{cor}
 Relations \eqref{rec-super3} and \eqref{rec-super4} are related to
\eqref{rec-super1} and \eqref{rec-super2} by the symmetry
\begin{equation*}\label{ZZmn}
Z^{n|m}\bigl(\mu(\bar x)|\mu(\bar t)\bigr) =
(-1)^{r_m}
\left. Z^{m|n}(\bar x|\bar t) \right|_{c \to -c} \\
\prod_{k=1}^{N-1}  f^{m|n}_{[k+1]}(\bar x^{k},  \bar x^{k+1}) f^{m|n}_{[k+1]}(\bar t^{k},  \bar t^{k+1})	,
\end{equation*}
with
\begin{equation}
\mu(\bar t)=\{\bar t^{m+n-1}+(n-1)c,\bar t^{n+m-2} +(n-2) c,\dots,\bar t^{m+1} + c,\bar t^{m}, \bar t^{m-1} + c, \dots, \bar t^{1}+(m-1)c\}.
\end{equation}
 One can also relate \eqref{rec-super3} to \eqref{rec-super4} using the symmetry \eqref{ZZmn-phi}.
Once more, the corollary \ref{cor-HC-susy} reduces to corollary \ref{cor-HC} when $m=0$ 
or $n=0$.
In the case $m=n=1$, the four recursions obtained from proposition \ref{Prop-HC-susy} and corollary \ref{cor-HC-susy} lead to the same equality $Z^{1|1}(\bar x|\bar t)=g(\bar x,\bar t)$, as expected for the highest coefficient of $\mathfrak{gl}(1|1)$-models, see \cite{HLPRS17a}. 


\section*{Conclusion}
In this paper we continued our study of Bethe vectors in $\mathfrak{gl}(m|n)$-invariant quantum integrable models. We developed a generalization of the zero mode method based on a twisting procedure. This twisted zero mode method allows us to deduce in a simple way
multiple action of all elements of the monodromy matrix on Bethe vectors.

In all cases, the multiple action is presented as a sum over partitions of sets of Bethe parameters.

Thanks to the multiple action formula, one can get an equivalent of the sum formula for the scalar product of off-shell Bethe vectors for $\mathfrak{gl}(m|n)$-invariant quantum integrable models \cite{HLPRS17a,Res86}. We provide new recursions for the highest coefficients entering the sum formula, the iteration being based on the rank of the algebra under consideration. In this way, the knowledge of the basic $\mathfrak{gl}(2)$, $\mathfrak{gl}(1|1)$ and $\mathfrak{gl}(2|1)$ models
 is enough to reconstruct the scalar product in the general case (at least theoretically).
The multiple action formula also allows to obtain form factors of the monodromy matrix elements, and we expect to report on it in a future work.

Of course, the sum formula \eqref{sp6} is not adapted to handle the thermodynamic limit of the models, and a determinant form for the scalar product and the form factors still needs to be found.
Yet, it is the first step in this direction, since it is (up to now) the only  general form known for the scalar product, up to few cases for $\mathfrak{gl}(2)$, $\mathfrak{gl}(3)$, $\mathfrak{gl}(1|1)$ and
$\mathfrak{gl}(2|1)$ models. However, let us  remark the Gaudin determinant formula for on-shell Bethe vectors, which has been established on general ground in \cite{HLPRS17b} for
$Y(\mathfrak{gl}(m|n))$ Yangians and in \cite{HLPRS18a}
for ${\cal U}_q(\widehat{\mathfrak{gl}}(n))$ quantum groups.

Finally, we want to stress that although the $RTT$ presentation is the framework used in this note,
our general strategy \cite{HLPRS17} is to use the current presentation of the Yangian.
Indeed, we believe that this current presentation is the right framework to achieve technical calculations in the higher rank models.
In particular, it allows to get the multiple action of  $T_{1,N+1}(z)$, which is the starting point of our twisted zero mode method.

Let us also note that the current presentation can be used for models based on
${\cal U}_q(\hat{\mathfrak{gl}}(m|n))$ deformed superalgebras. Results for these models will be presented elsewhere.

\section*{Acknowledgments}
The work of S.P. was supported in part by the RFBR grant 19-01-00726-a.
Part of this work was done during the visit of A.L. and N.S. to LAPTh, 
with financial support from the USMB grant AAP ASI-32.
The work A. L. was carried out in Skolkovo Institute of Science and Technology   under financial support of Russian Science Foundation within grant 19-11-00275. 
The work of A.H. was supported by the BME-Nanotechnology FIKP grant of ITM (BME FIKP-NAT).

\appendix

\section{Proof of lemma~\ref{lem1}\label{ApA}}

We use induction to prove lemma~\ref{lem1}.
First of all, we observe that for $i=1$ and $j=N+1$,  \r{ac222}
reduces to \r{ac3}, so that the action formula for $T_{1,N+1}(z)$ is proven.
Next, let us assume that equation \r{ac222} is valid for some values
$(i,j)$. We consider two particular cases of \r{zm}:
\begin{eqnarray}\label{zm1}
[T_{i,j}(z),T_{i+1,i}[0]] &=& \kappa_i\ T_{i+1,j}(z)  - \delta_{i,j-1}\ \kappa_{i+1}\   T_{i,i}(z),
\\
\label{zm2}
[T_{i,j}(z),T_{j,j-1}[0]] &=& - \kappa_j\   T_{i,j-1}(z) + \delta_{i,j-1}\ \kappa_{j-1}\ T_{j,j}(z)
\end{eqnarray}
When considering these commutation relations, we can separately equate the coefficients
for different twisting parameters, since $\kappa_i$ are arbitrary complex numbers.
Then, from \eqref{zm1},
 we can derive  the action formula for the elements $T_{i+1,j}(z)$ and $T_{i,i}(z)$.
 Similarly, using \eqref{zm2},
we obtain the action of the elements $T_{i,j-1}(z)$ and $T_{j,j}(z)$.
We examine each of these actions separately. The action formula for the elements
$T_{i+1,j}(z)$ (resp. $T_{i,j-1}(z)$) will provide a recursion for $i$ (resp. $j$).
The actions of $T_{i,i}(z)$ and $T_{j,j}(z)$ give consistency checks of the formulas.

\subsection{Action of $T_{i+1,j}(z)$}
It corresponds to the terms proportional to $\kappa_i$ in \eqref{zm1}.
We split this calculation in two cases, corresponding to the relative position of $i$ and $j$.

\paragraph{Case 1: $\boldsymbol{j\geq i+1}$.}   Then,
the last factor in first line of  \r{ac222} disappears and this action formula becomes
\begin{equation}\label{ac5}
T_{i,j}(z) \mathbb{B}(\bar t) =  \lambda_{N+1}(z)
      \sum_{{\rm part}} \mathbb{B}(\bar w_{\st} )
        \prod_{s=1}^{i-1} \frac{ f(\bar w^s_{\so},\bar w^s_{\st})}
        {h(\bar w^s_{\so}, \bar w^{s-1}_{\so}) f(\bar w^s_{\so},\bar w^{s-1}_{\st})}
		\prod_{s=j}^{N}      \frac{\alpha_{s}(\bar w^{s}_{\sth})   f(\bar w^s_{\st},\bar w^s_{\sth})   }
		{h(\bar w^{s+1}_{\sth}, \bar w^{s}_{\sth})f(\bar w^{s+1}_{\st}, \bar w^{s}_{\sth})}.
\end{equation}
 Here the sets $\bar w^s=\{\bar t^s,z\}$ for $s=i,\ldots,j-1$ are not divided into subsets,
 and we have $\bar w^s_{\st}=\bar w^s$ and
$\bar w^s_{\so}=\bar w^s_{\sth}=\varnothing$. Using formulas \r{ac4} and \r{ac5} we find
\begin{equation}\label{ac6}
\begin{split}
     & T_{i,j}(z) T_{i+1,i}[0] \mathbb{B}(\bar t)\Big|_{\kappa_{i+1}=0} = -\kappa_i \lambda_{N+1}(z)
      \sum_{\ell=1}^{r_i}\frac{f(t^i_\ell,\bar t^i_\ell)}{f(t^i_\ell,\bar t^{i-1})}
      \sum_{{\rm part}} \mathbb{B}(\bar w_{\st} )
        \\
   &\quad\times    \prod_{s=1}^{i-1} \frac{ f(\bar w^s_{\so},\bar w^s_{\st})}
   {h(\bar w^s_{\so}, \bar w^{s-1}_{\so}) f(\bar w^s_{\so},\bar w^{s-1}_{\st})}
		\prod_{s=j}^{N}      \frac{\alpha_{s}(\bar w^{s}_{\sth})   f(\bar w^s_{\st},\bar w^s_{\sth})   }
		{h(\bar w^{s+1}_{\sth}, \bar w^{s}_{\sth})f(\bar w^{s+1}_{\st}, \bar w^{s}_{\sth})},
		      \end{split}
    \end{equation}
    where $\bar w^i_{\st}=\{\bar t^i_\ell,z\}$. On the other hand,
    the action of the same operators in the reverse order can be written as
\begin{equation}\label{ac7}
\begin{split}
     & T_{i+1,i}[0] T_{i,j}(z)  \mathbb{B}(\bar t)\Big|_{\kappa_{i+1}=0} = -\kappa_i \lambda_{N+1}(z)
      \sum_{{\rm part}} \mathbb{B}(\bar w_{\st} )
        \\
   &\quad\times \frac{f(\bar w^i_{\so},\bar w^i_{\st})}{f(\bar w^i_{\so},\bar w^{i-1}_{\st})}
      \prod_{s=1}^{i-1} \frac{ f(\bar w^s_{\so},\bar w^s_{\st})}
   {h(\bar w^s_{\so}, \bar w^{s-1}_{\so}) f(\bar w^s_{\so},\bar w^{s-1}_{\st})}
		\prod_{s=j}^{N}      \frac{\alpha_{s}(\bar w^{s}_{\sth})   f(\bar w^s_{\st},\bar w^s_{\sth})   }
		{h(\bar w^{s+1}_{\sth}, \bar w^{s}_{\sth})f(\bar w^{s+1}_{\st}, \bar w^{s}_{\sth})}.
		      \end{split}
    \end{equation}
    Here the set $\bar w^i=\{\bar t^i,z\}$ is divided into subsets
    $\{\bar w^i_{\so},\bar w^i_{\st}\}\vdash \bar w^i$ such that $\# \bar w^i_{\so}=1$.
    The sum over
    $\ell$ in the first line of \r{ac6} can also be presented as the sum over these partitions.
    To do this, we transform the ratio
    \begin{equation}\label{ac8}
    \frac{f(t^i_\ell,\bar t^i_\ell)}{f(t^i_\ell,\bar t^{i-1})}=\frac{f(t^i_\ell,\bar w^i_{\st})}{f(t^i_\ell,\bar w^{i-1})},
    \end{equation}
where $\bar w^i_{\st}=\{\bar t^i_\ell,z\}$, and add to the sum over $\ell$
one zero term proportional to
\begin{equation*}
\frac{f(z,\bar t^i)}{f(z,\bar w^{i-1})}=0.
\end{equation*}
Then the sum over $\ell$ takes the form
\begin{equation}\label{ac9}
 \sum_{\ell=1}^{r_i}\frac{f(t^i_\ell,\bar t^i_\ell)}{f(t^i_\ell,\bar t^{i-1})}(\ \cdot\ )=
 \sum_{{\rm part}} \frac{f(\bar w^i_{\so},\bar w^i_{\st})}{f(\bar w^i_{\so},\bar w^{i-1}_{\st})}
 \frac{1}{f(\bar w^i_{\so},\bar w^{i-1}_{\so})}(\ \cdot\ ),
\end{equation}
where the latter sum runs over partitions
$\{\bar w^i_{\so},\bar w^i_{\st}\}\vdash \bar w^i$ such that $\# \bar w^i_{\so}=1$.
Subtracting now \r{ac6} and \r{ac7} and using a trivial identity
\begin{equation}\label{ac10}
\left(1- \frac{1} {f(\bar w^i_{\so},\bar w^{i-1}_{\so})}\right)= \frac{1}{h(\bar w^i_{\so},\bar w^{i-1}_{\so})},
\end{equation}
we obtain from \r{zm1} that
\begin{equation}\label{ac11}
T_{i+1,j}(z) \mathbb{B}(\bar t) =  \lambda_{N+1}(z)
      \sum_{{\rm part}} \mathbb{B}(\bar w_{\st} )
        \prod_{s=1}^{i} \frac{ f(\bar w^s_{\so},\bar w^s_{\st})}
        {h(\bar w^s_{\so}, \bar w^{s-1}_{\so}) f(\bar w^s_{\so},\bar w^{s-1}_{\st})}
		\prod_{s=j}^{N}      \frac{\alpha_{s}(\bar w^{s}_{\sth})   f(\bar w^s_{\st},\bar w^s_{\sth})   }
		{h(\bar w^{s+1}_{\sth}, \bar w^{s}_{\sth})f(\bar w^{s+1}_{\st}, \bar w^{s}_{\sth})}.
\end{equation}

\paragraph{Case 2: $\boldsymbol{j<i+1}$.}
Let us repeat the calculations above for the case $i\geq j$. Instead of the formula \r{ac6} we may write
\begin{equation}\label{ac12}
\begin{split}
     & T_{i,j}(z) T_{i+1,i}[0] \mathbb{B}(\bar t)\Big|_{\kappa_{i+1}=0} = -\kappa_i \lambda_{N+1}(z)
      \sum_{\ell=1}^{r_i}\frac{f(t^i_\ell,\bar t^i_\ell)}{f(t^i_\ell,\bar t^{i-1})}
      \sum_{{\rm part}} \mathbb{B}(\bar w_{\st} )
      \frac{\prod_{s=j}^{i-1} f(\bar w^s_{\so},\bar w^s_{\sth})} {\prod_{s=j}^{i-2} f(\bar w^{s+1}_{\so},\bar w^s_{\sth}) }
        \\
   &\qquad\qquad\times    \prod_{s=1}^{i-1} \frac{ f(\bar w^s_{\so},\bar w^s_{\st})}
   {h(\bar w^s_{\so}, \bar w^{s-1}_{\so}) f(\bar w^s_{\so},\bar w^{s-1}_{\st})}
		\prod_{s=j}^{N}      \frac{\alpha_{s}(\bar w^{s}_{\sth})   f(\bar w^s_{\st},\bar w^s_{\sth})   }
		{h(\bar w^{s+1}_{\sth}, \bar w^{s}_{\sth})f(\bar w^{s+1}_{\st}, \bar w^{s}_{\sth})},
		      \end{split}
    \end{equation}
    where now the set $\bar w^i_{\st}$ is obtained by the partition
 $\{\bar w^i_{\st}, \bar w^i_{\sth}\}\vdash \{\bar t^i_\ell,z\}$ such that $\# \bar w^i_{\sth}=1$.
 Taking into account that $\{\bar w^{i-1}_{\so},\bar w^{i-1}_{\st}, \bar w^{i-1}_{\sth}\}\vdash \bar w^{i-1}$
 and transforming the ratio
 \begin{equation}\label{ac13}
 \frac{f(t^i_\ell,\bar t^i_\ell)}{f(t^i_\ell,\bar t^{i-1})}=\frac{f(t^i_\ell,\bar w^i_{\st})f(t^i_\ell, \bar w^i_{\sth})}
 {f(t^i_\ell,\bar w^{i-1})}=\frac{f(\bar w^i_{\so},\bar w^i_{\st})f(\bar w^i_{\so},\bar w^i_{\sth})}
 {f(\bar w^i_{\so}, \bar w^{i-1}_{\so})f(\bar w^i_{\so},\bar w^{i-1}_{\st})f(\bar w^i_{\so}, \bar w^{i-1}_{\sth})},
 \end{equation}
 we can rewrite \r{ac12} as follows:
 \begin{equation}\label{ac14}
\begin{split}
     & T_{i,j}(z) T_{i+1,i}[0] \mathbb{B}(\bar t)\Big|_{\kappa_{i+1}=0} = -\kappa_i \lambda_{N+1}(z)
      \sum_{{\rm part}} \mathbb{B}(\bar w_{\st} )
      \frac{\prod_{s=j}^{i} f(\bar w^s_{\so},\bar w^s_{\sth})} {\prod_{s=j}^{i-1} f(\bar w^{s+1}_{\so},\bar w^s_{\sth}) }
        \\
   &\qquad\qquad\times  \frac{1}{g(\bar w^i_{\so}, \bar w^{i-1}_{\so})}  \prod_{s=1}^{i} \frac{ f(\bar w^s_{\so},\bar w^s_{\st})}
   {h(\bar w^s_{\so}, \bar w^{s-1}_{\so}) f(\bar w^s_{\so},\bar w^{s-1}_{\st})}
		\prod_{s=j}^{N}      \frac{\alpha_{s}(\bar w^{s}_{\sth})   f(\bar w^s_{\st},\bar w^s_{\sth})   }
		{h(\bar w^{s+1}_{\sth}, \bar w^{s}_{\sth})f(\bar w^{s+1}_{\st}, \bar w^{s}_{\sth})}.
		      \end{split}
    \end{equation}

 When we act by the operators in reverse order, we first use the induction assumption
 \r{ac222}. According to this assumption,
 sets $\bar w^s$ are divided into $\{\bar w^s_{\so},\bar w^s_{\st},\bar w^s_{\sth}\}
 \vdash \bar w^s$, but for the specific
   set $\bar w^i$ we use the temporary notation $\bar w^i_{\sii}$ instead of $\bar w^i_{\st}$, so that
   it is divided into $\{\bar w^i_{\sii},\bar w^i_{\sth}\}
 \vdash \bar w^i$
 (recall that  $\bar w^i_{\so}=\varnothing$ due to the $(i,j)$-condition, see definition \ref{def:ij-cond}).
  Let us rewrite the
 induction assumption \r{ac222} in the form
 \begin{equation}\label{ac2bb}
\begin{split}
     & T_{i,j}(z) \mathbb{B}(\bar t) =  \lambda_{N+1}(z)
      \sum_{{\rm part}} \mathbb{B}(\bar w^1_{\st},\ldots, \bar w^{i-1}_{\st},\bar w^i_{\sii},
      \bar w^{i+1}_{\st},\ldots,\bar w^N_{\st} )
     \frac{\prod_{s=j}^{i-1} f(\bar w^s_{\so},\bar w^s_{\sth})} {\prod_{s=j}^{i-2} f(\bar w^{s+1}_{\so},\bar w^s_{\sth}) } \\
     &\quad \times
 \prod_{s=1}^{i-1} \frac{ f(\bar w^s_{\so},\bar w^s_{\st})}
   {h(\bar w^s_{\so}, \bar w^{s-1}_{\so}) f(\bar w^s_{\so},\bar w^{s-1}_{\st})}
    \prod_{{s=j\atop s\neq i,i-1}}^{N}
 \frac{\alpha_{s}(\bar w^{s}_{\sth})
 f(\bar w^s_{\st},\bar w^s_{\sth}) }
{h(\bar w^{s+1}_{\sth}, \bar w^{s}_{\sth})f(\bar w^{s+1}_{\st}, \bar w^{s}_{\sth})}
\\
&\quad\times  \frac{f(\bar w^i_{\sii},\bar w^i_{\sth})}{f(\bar w^{i}_{\sii}, \bar w^{i-1}_{\sth})}
 \frac{\alpha_{i-1}(\bar w^{i-1}_{\sth}) f(\bar w^{i-1}_{\st},\bar w^{i-1}_{\sth})}
{h(\bar w^{i}_{\sth}, \bar w^{i-1}_{\sth})}\
 \frac{\alpha_{i}(\bar w^{i}_{\sth}) }
{h(\bar w^{i+1}_{\sth}, \bar w^{i}_{\sth})f(\bar w^{i+1}_{\st}, \bar w^{i}_{\sth})}.
\end{split}
\end{equation}
where we have singled out the terms where the subset $\bar w^i_{\sii}$ occurs
(last line of the equation). Note that the  resulting
 Bethe vector also depends on the auxiliary subset $\bar w^i_{\sii}$, as it is shown explicitly in \r{ac2bb}.

Now we apply the zero mode operator $T_{i+1,i}[0]$ on both sides of \r{ac2bb} and take into account
only the part proportional to $\kappa_i$. This action
divides the auxiliary subset $\bar w^i_{\sii}$ into $\{\bar w^i_{\so},\bar w^i_{\st}\}\vdash \bar w^i_{\sii}$ and
produces an additional factor $-\frac{f(\bar w^i_{\so},\bar w^i_{\st})}{f(\bar w^i_{\so},\bar w^{i-1}_{\st})}$.
This factor together with the first one
$ \frac{f(\bar w^i_{\sii},\bar w^i_{\sth})}{f(\bar w^{i}_{\sii}, \bar w^{i-1}_{\sth})}$
in the last line of \r{ac2bb} may be factorized as follows
\begin{equation}\label{cl1}
-\frac{f(\bar w^i_{\so},\bar w^i_{\st})}{f(\bar w^i_{\so},\bar w^{i-1}_{\st})}
\frac{f(\bar w^i_{\sii},\bar w^i_{\sth})}{f(\bar w^{i}_{\sii}, \bar w^{i-1}_{\sth})}
=
-\frac{f(\bar w^i_{\so},\bar w^i_{\st})}{f(\bar w^i_{\so},\bar w^{i-1}_{\st})}
\frac{f(\bar w^i_{\so},\bar w^i_{\sth})}{f(\bar w^{i}_{\so}, \bar w^{i-1}_{\sth})}
\frac{f(\bar w^i_{\st},\bar w^i_{\sth})}{f(\bar w^{i}_{\st}, \bar w^{i-1}_{\sth})}.
\end{equation}

The first factor $\frac{f(\bar w^i_{\so},\bar w^i_{\st})}{f(\bar w^i_{\so},\bar w^{i-1}_{\st})}$
    in the right hand side of \r{cl1} will shift index $i-1\to i$ in the first product of the second line in \r{ac2bb}
    producing the factor $h(\bar w^i_{\so},\bar w^{i-1}_{\so})$. The second factor
 $ \frac{f(\bar w^i_{\so},\bar w^i_{\sth})}{f(\bar w^{i}_{\so}, \bar w^{i-1}_{\sth})}$ from r.h.s. of \r{cl1}
 change index $i\to i+1$ in the product of the first line of \r{ac2bb}. Finally, the last
 factor $\frac{f(\bar w^i_{\st},\bar w^i_{\sth})}{f(\bar w^{i}_{\st}, \bar w^{i-1}_{\sth})}$
 allows to restore the values $s=i-1$ and $s=i$ in the product from $j$ to $N$ of the
 second line of \r{ac2bb}.

Thus for the consecutive action of the operators $T_{i+1,i}[0]$ and  $T_{i,j}(z)$
onto the off-shell Bethe vector $\mathbb{B}(\bar t)$ in the case $i\geq j$ we  get
 \begin{equation}\label{ac20}
\begin{split}
     & T_{i+1,i}[0] T_{i,j}(z)  \mathbb{B}(\bar t)\Big|_{\kappa_{i+1}=0} = -\kappa_i \lambda_{N+1}(z)
      \sum_{{\rm part}} \mathbb{B}(\bar w_{\st} )
      \frac{\prod_{s=j}^{i} f(\bar w^s_{\so},\bar w^s_{\sth})} {\prod_{s=j}^{i-1} f(\bar w^{s+1}_{\so},\bar w^s_{\sth}) }
        \\
   &\quad\times h(\bar w^i_{\so},\bar w^{i-1}_{\so})
      \prod_{s=1}^{i} \frac{ f(\bar w^s_{\so},\bar w^s_{\st})}
   {h(\bar w^s_{\so}, \bar w^{s-1}_{\so}) f(\bar w^s_{\so},\bar w^{s-1}_{\st})}
		\prod_{s=j}^{N}      \frac{\alpha_{s}(\bar w^{s}_{\sth})   f(\bar w^s_{\st},\bar w^s_{\sth})   }
		{h(\bar w^{s+1}_{\sth}, \bar w^{s}_{\sth})f(\bar w^{s+1}_{\st}, \bar w^{s}_{\sth})}.
		      \end{split}
    \end{equation}

 Subtracting \r{ac20} from \r{ac14} and using \r{zm1} we finally prove from the
 identity $h(x,y)-g(x,y)^{-1}=1$ that  the action of the matrix element $T_{i+1,j}(z)$ is given by
 the formula \r{ac2} at the shifted index $i\to i+1$.

\paragraph{Recursion on $\boldsymbol i$.}
The two above cases show by induction that the action formula \eqref{ac2} is valid for $T_{i+1,j}(z)$, provided the terms proportional to $\kappa_{i+1}$ add up correctly. This is showed in section \ref{sec:Tii}. Then, the induction shows that the action formula is valid for $T_{k,j}(z)$, $\forall k\geq i$,
provided \eqref{ac2} is valid for $T_{i,j}(z)$.

In particular,
since we know it is valid for $T_{1,N+1}(z)$, we know that the action formula is valid for
 $T_{k,N+1}(z)$, $\forall k$.

\subsection{Action of $T_{i,i}(z)$\label{sec:Tii}}
Now we have to check that
the terms coming from the action of the $[T_{i,j}(z),T_{i+1,i}[0]]$  onto $\mathbb{B}(\bar t)$
and proportional to the twisting parameter $\kappa_{i+1}$
cancel each other for $i\neq j-1$ and produce the action of the operator
$T_{i,i}(z)$ for $i=j-1$. We split this study in three cases.

\paragraph{Case 1: $\boldsymbol{j>i+1}$.} Indeed,  for $i<j-1$
the terms proportional to $\kappa_{i+1}$ after the action of the operators   $T_{i,j}(z) T_{i+1,i}[0]$
onto off-shell Bethe vector  $ \mathbb{B}(\bar t)$ are
\begin{equation}\label{ac66}
\begin{split}
      T_{i,j}(z) T_{i+1,i}[0] \mathbb{B}(\bar t)\Big|_{\kappa_{i}=0} &= \kappa_{i+1} \lambda_{N+1}(z)
      \sum_{\ell=1}^{r_i}\frac{\alpha_i(t^i_\ell)f(\bar t^i_\ell,t^i_\ell)}{f(\bar t^{i+1},t^{i}_\ell)}
      \sum_{{\rm part}} \mathbb{B}(\bar w_{\st} )
        \\
   &\quad\times    \prod_{s=1}^{i-1} \frac{ f(\bar w^s_{\so},\bar w^s_{\st})}
   {h(\bar w^s_{\so}, \bar w^{s-1}_{\so}) f(\bar w^s_{\so},\bar w^{s-1}_{\st})}
		\prod_{s=j}^{N}      \frac{\alpha_{s}(\bar w^{s}_{\sth})   f(\bar w^s_{\st},\bar w^s_{\sth})   }
		{h(\bar w^{s+1}_{\sth}, \bar w^{s}_{\sth})f(\bar w^{s+1}_{\st}, \bar w^{s}_{\sth})}.
		      \end{split}
    \end{equation}

We again present the sum over $\ell$  as sum over partitions of the
set $\{\bar w^i_{\st},\bar w^i_{\sth}\}\vdash \bar w^i=\{\bar t^i,z\}$ with $\#\bar w^i_{\sth}=1$:
\begin{equation}\label{ac66bis}
\begin{split}
      T_{i,j}(z) T_{i+1,i}[0] \mathbb{B}(\bar t)\Big|_{\kappa_{i}=0} &= \kappa_{i+1} \lambda_{N+1}(z)
      \sum_{{\rm part}} \mathbb{B}(\bar w_{\st} )\frac{\alpha_i(\bar w^i_{\sth})f(\bar w^i_{\st},\bar w^i_{\sth})}{f(\bar w^{i+1},\bar w^{i}_{\sth})}
        \\
   &\quad\times    \prod_{s=1}^{i-1} \frac{ f(\bar w^s_{\so},\bar w^s_{\st})}
   {h(\bar w^s_{\so}, \bar w^{s-1}_{\so}) f(\bar w^s_{\so},\bar w^{s-1}_{\st})}
		\prod_{s=j}^{N}      \frac{\alpha_{s}(\bar w^{s}_{\sth})   f(\bar w^s_{\st},\bar w^s_{\sth})   }
		{h(\bar w^{s+1}_{\sth}, \bar w^{s}_{\sth})f(\bar w^{s+1}_{\st}, \bar w^{s}_{\sth})}.
		      \end{split}
    \end{equation}
The additional term in \eqref{ac66bis} (w.r.t. \eqref{ac66}), corresponds to $\bar w^i_{\sth}=\{z\}$
and is in fact zero, due to the factor
$f(\bar w^{i+1},z)^{-1}$ in the first line of \eqref{ac66bis}.

To present the action in the reverse order, we define $\bar y^i=\bar w^i_{\st}$:
 \begin{equation}\label{ac66ter}
\begin{split}
& T_{i+1,i}[0]  T_{i,j}(\bar z) \mathbb{B}(\bar t)\Big|_{\kappa_{i}=0} = \kappa_{i+1} \lambda_{N+1}(\bar z)
 \sum_{{\rm part}}  \sum_{\ell=1}^{r_i}
\frac{\alpha_i( y^i_\ell)f(\bar y^i_\ell, y^i_\ell)}{f(\bar w^{i+1}_{\st},y^i_\ell)}
\mathbb{B}(\bar w_{\st}\setminus\{y^i_\ell\} )
\\
   &\quad\times
 \prod_{s=1}^{i-1} \frac{ f(\bar w^s_{\so},\bar w^s_{\st})}
   {h(\bar w^s_{\so}, \bar w^{s-1}_{\so}) f(\bar w^s_{\so},\bar w^{s-1}_{\st})}
   \prod_{s=j}^{N}
 \frac{\alpha_{s}(\bar w^{s}_{\sth})
 f(\bar w^s_{\st},\bar w^s_{\sth}) }
{h(\bar w^{s+1}_{\sth}, \bar w^{s}_{\sth})f(\bar w^{s+1}_{\st}, \bar w^{s}_{\sth})}.
\end{split}
\end{equation}
Once more, we can transform the sum on $\ell$ as a sum over partitions. Indeed,
the same factor $f(\bar w^{i+1},z)^{-1}$ appears,
since $\bar w^{i+1}_{\st}\equiv \bar w^{i+1}$.
Thus, \eqref{ac66ter} becomes identical to \eqref{ac66bis}, and we get 0 as a final result, as expected from \eqref{zm1}.

\paragraph{Case 2: $\boldsymbol{j=i+1}$.}   In this case, the set $\bar w^{i+1}$ in the action
of the element $T_{i,i+1}(z)$ is divided into subsets $\bar w^{i+1}_{\st}$ and $\bar w^{i+1}_{\sth}$. Then
the actions of  $T_{i,j}(z) T_{i+1,i}[0] \mathbb{B}(\bar t)$ and
$T_{i+1,i}[0] T_{i,j}(z)  \mathbb{B}(\bar t)$ at $\kappa_i=0$ give different results.
 The first action is given by
\r{ac66bis} while for the reverse action the last factor of the first line is replaced by
\begin{equation*}
\frac{\alpha_i(\bar w^i_{\sth})f(\bar w^i_{\st},\bar w^i_{\sth})}{f(\bar w^{i+1},\bar w^{i}_{\sth})}
\quad\to\quad
\frac{\alpha_i(\bar w^i_{\sth})f(\bar w^i_{\st},\bar w^i_{\sth})}{f(\bar w^{i+1}_{\st},\bar w^{i}_{\sth})}.
\end{equation*}
Due to \r{ac10}, the difference of these actions  produces the action of the monodromy matrix element
$T_{i,i}(z)$ onto $\mathbb{B}(\bar t)$, which corresponds to the second term
in the right hand side of \r{zm1}.

\paragraph{Case 3: $\boldsymbol{i+1>j}$.}  Again, to conclude the induction proof for the case
 $i\geq j$ we have to verify that the terms at the twisting parameter $\kappa_{i+1}$  cancel each other in the
 action formulas $T_{i,j}(z) T_{i+1,i}[0]   \mathbb{B}(\bar t)$ and $T_{i+1,i}[0] T_{i,j}(z)  \mathbb{B}(\bar t)$.
 We leave this exercise to the interested reader.

\subsection{Action of $T_{i,j-1}(z)$ and end of the recursion}
In exactly the same way, we can prove the validity of the action formula \r{ac2} for $T_{i,j-1}(z)$ if it is valid for $T_{i,j}(z)$. This is done
starting with the action \r{ac3} and using the commutation relation \r{zm2}, together with the splitting over the twisting parameters $\kappa_j$ and $\kappa_{j-1}$. Induction then proves that it is valid for $T_{i,k}(z)$, $\forall k\leq j$.

Finally, since the induction on $i$ showed that the action formula  \r{ac222} is valid for
 $T_{k,N+1}(z)$, $\forall k$, the induction on $j$ proves it is valid for $T_{k,\ell}(z)$, $\forall k,\ell$.
\qed

\section{Proof of proposition~\ref{prop1}\label{ApB}}

In the previous appendix, we have proved lemma \ref{lem1}. Since it corresponds to equation \r{ac2} for $p=1$, it is the base of the induction on $p$ that we are using to prove the general case.

 For simplicity, we  first consider the case  $i<j$. We assume that \r{ac2} is valid for
the cardinality  $\#\bar z=p-1$ of the set $\bar z$.
Then, for $\#\bar z=p$, we can apply \r{ac2} for the successive action of $T_{i,j}(z_1)$
and $T_{i,j}(\bar z_1)$  to get
\begin{equation}\label{ac21}
\begin{split}
     & T_{i,j}(z_1)T_{i,j}(\bar z_1) \mathbb{B}(\bar t) =  \lambda_{N+1}(\bar z_1)
      \sum_{{\rm part}} T_{i,j}(z_1) \mathbb{B}(\bar w_{\sii} )     \\
   &\qquad\times    \prod_{s=1}^{i-1} \frac{K(\bar w^s_{\si}| \bar w^{s-1}_{\si}) f(\bar w^s_{\si},\bar w^s_{\sii})}
   {f(\bar w^s_{\si}, \bar w^{s-1}_{\si}) f(\bar w^s_{\si},\bar w^{s-1}_{\sii})}
		\prod_{s=j}^{N}      \frac{\alpha_{s}(\bar w^{s}_{\siii})
		K(\bar w^{s+1}_{\siii}|\bar w^{s}_{\siii})  f(\bar w^s_{\sii},\bar w^s_{\siii})   }
		{f(\bar w^{s+1}_{\siii}, \bar w^{s}_{\siii})f(\bar w^{s+1}_{\sii}, \bar w^{s}_{\siii})}       \\
 &\quad= \lambda_{N+1}(\bar z)  \sum_{{\rm part}}  \mathbb{B}(\bar w_{\st} )     \\
 &\qquad\times    \prod_{s=1}^{i-1} \frac{K(\bar w^s_{\si}| \bar w^{s-1}_{\si}) f(\bar w^s_{\si},\bar w^s_{\sii})}
   {f(\bar w^s_{\si}, \bar w^{s-1}_{\si}) f(\bar w^s_{\si},\bar w^{s-1}_{\sii})}
		\prod_{s=j}^{N}      \frac{\alpha_{s}(\bar w^{s}_{\siii})
		K(\bar w^{s+1}_{\siii}|\bar w^{s}_{\siii})  f(\bar w^s_{\sii},\bar w^s_{\siii})   }
		{f(\bar w^{s+1}_{\siii}, \bar w^{s}_{\siii})f(\bar w^{s+1}_{\sii},\bar  w^{s}_{\siii})}       \\
		&\qquad\times    \prod_{s=1}^{i-1} \frac{K(\bar w^s_{\siv}| \bar w^{s-1}_{\siv})
		f(\bar w^s_{\siv},\bar w^s_{\st})}
   {f(\bar w^s_{\siv}, \bar w^{s-1}_{\siv}) f(\bar w^s_{\siv},\bar w^{s-1}_{\st})}
		\prod_{s=j}^{N}      \frac{\alpha_{s}(\bar w^{s}_{\sv}) K(\bar w^{s+1}_{\sv}|\bar w^{s}_{\sv})
		f(\bar w^s_{\st},\bar w^s_{\sv})   }
		{f(\bar w^{s+1}_{\sv}, \bar w^{s}_{\sv})f(\bar w^{s+1}_{\st}, \bar w^{s}_{\sv})}.
		      \end{split}
    \end{equation}
Here\footnote{We remind that by convention $\bar w^0 = \bar w^{N+1} =
\bar z$.} for  $0 \leq s<i$, the sums in \r{ac21} run  over partitions  $\{\bar w^s_{\si},\bar w^s_{\sii}\}\vdash \bar w^s$
with cardinality $\#\bar w^s_{\si}=p-1$, and then
over partition $\{\bar w^s_{\siv},\bar w^s_{\st}\}\vdash \bar w^s_{\sii}$ with cardinality
$\#\bar w^s_{\siv}=1$.  Similarly, for $j\leq s\leq N+1$, the sums in \r{ac21} run
over partitions  $\{\bar w^s_{\sii},\bar w^s_{\siii}\}\vdash \bar w^s$
with cardinality $\#\bar w^s_{\siii}=p-1$ and then
over partition $\{\bar w^s_{\sv},\bar w^s_{\st}\}\vdash \bar w^s_{\sii}$ with cardinality
$\#\bar w^s_{\sv}=1$. Note that for $i\leq s<j$, the subsets $\bar w^s_{\sii}$ and $\bar w^s_{\st}$ are equal to $\bar w^s$. Thus, all other sets
$\bar w^s_{\si}$, $\bar w^s_{\siii}$, $\bar w^s_{\siv}$, and $\bar w^s_{\sv}$  
are empty in that case.

Using properties of the Izergin determinant we can combine the sets $\bar w^s_{\si}\cup \bar w^s_{\siv}=\bar w^s_{\so}$ for $s<i$ (resp.
 $\bar w^s_{\siii}\cup \bar w^s_{\sv}=\bar w^s_{\sth}$ for $s\geq j$)
 of  cardinalities
$\#\bar w^s_{\so}=p$ (resp. $\#\bar w^s_{\sth}=p)$ and rewrite \r{ac21} as sums over partitions
$\{\bar w^s_{\so},\bar w^s_{\st}\}\vdash \bar w^s$ for $0\leq s<i$ and sums over partitions
 $\{\bar w^s_{\st},\bar w^s_{\sth}\}\vdash \bar w^s$ for $j\leq s\leq N+1$.
  Indeed,  we  may factorize the ratio in the fourth line of \r{ac21}
 \begin{equation*}
 \frac{f(\bar w^s_{\si},\bar w^s_{\sii})} {f(\bar w^s_{\si},\bar w^{s-1}_{\sii})}=
  \frac{f(\bar w^s_{\si},\bar w^s_{\siv})} {f(\bar w^s_{\si},\bar w^{s-1}_{\siv})}
   \frac{f(\bar w^s_{\si},\bar w^s_{\st})} {f(\bar w^s_{\si},\bar w^{s-1}_{\st})}
 \end{equation*}
 and writing explicitly all the factors depending on the sets $\bar w^s_{\si}$ and $\bar w^s_{\siv}$
 for any fixed $s$ from the interval
 $[1,\ldots,i-1]$ we obtain the sum
 \begin{equation}\label{ac22}
 \begin{split}
& \sum_{\{\bar w^s_{\si},\bar w^s_{\siv}\}\vdash \bar w^s_{\so}}
 \frac{f(\bar w^s_{\si},\bar w^{s}_{\siv})}{f(\bar w^s_{\si},\bar w^{s-1}_{\siv})}
 \frac{ K(\bar w^s_{\siv}| \bar w^{s-1}_{\siv})}{f(\bar w^s_{\siv}, \bar w^{s-1}_{\siv})}
 \frac{K(\bar w^s_{\si}| \bar w^{s-1}_{\si})}{f(\bar w^s_{\si}, \bar w^{s-1}_{\si})}=\\
 &\quad = \frac{(-1)^{p-1}}{f(\bar w^s_{\so},\bar w^{s-1}_{\siv})}
 \sum_{\{\bar w^s_{\si},\bar w^s_{\siv}\}\vdash \bar w^s_{\so}}
 f(\bar w^s_{\si},\bar w^{s}_{\siv}) K(\bar w^s_{\siv}| \bar w^{s-1}_{\siv})
 K(\bar w^{s-1}_{\si}-c| \bar w^{s}_{\si})=\\
 &\quad = (-1)^p K(\{\bar w^{s-1}_{\siv}-c,\bar w^{s-1}_{\si}-c\}|\bar w^s_{\so})=
 (-1)^p K(\bar w^{s-1}_{\so}-c|\bar w^s_{\so})=
 \frac{K(\bar w^s_{\so}|\bar w^{s-1}_{\so})}{f(\bar w^s_{\so}|\bar w^{s-1}_{\so})}.
 \end{split}
 \end{equation}
Here we used the following property of the Izergin determinant
\begin{equation}\label{Izp}
K(\bar x-c|\bar y)=(-1)^p\frac{K(\bar y|\bar x)}{f(\bar y|\bar x)},\qquad \text{for }\ \#\bar x=p,
\end{equation}
and a summation identity  \cite{BelPRS12a}
  \begin{equation}\label{LM1}
	\sum_{\{\bar w_{\so},\bar w_{\st}\}\vdash \bar w}
      K(\bar w_{\so}|\bar u)K(\bar v|\bar w_{\st})f(\bar w_{\st},\bar w_{\so})
      = (-1)^{m_1}f(\bar w,\bar u) K(\{\bar u-c,\bar v\}|\bar w).
      \end{equation}
 Here $\bu$, $\bv$, and $\bw$ are sets of arbitrary complex numbers such that $\#\bu=m_1$, $\#\bv=m_2$, and $\#\bw=m_1+m_2$.
      The sum in \r{LM1} is taken with respect to all partitions of the set $\bar w$ into
      subsets $\bar w_{\so}$ and $\bar w_{\st}$ with $\#\bar w_{\so}=m_1$ and $\#\bar w_{\st}=m_2$.

To apply \r{LM1} to \r{ac22} we identify:
$\bar w_{\so}=\bar w^s_{\siv}$, $\bar w_{\st}=\bar w^s_{\si}$, $\bar u=\bar w^{s-1}_{\siv}$,
$\bar v=\bar w^{s-1}_{\si}-c$, $m_1=1$ and $m_2=p-1$. Similarly using \r{Izp} and \r{LM1}
we find that for $j\leq s\leq N$
\begin{equation}\label{ac23}
 \sum_{\{\bar w^s_{\siii},\bar w^s_{\sv}\}\vdash \bar w^s_{\sth}}
 \frac{f(\bar w^s_{\sv},\bar w^{s}_{\siii})}{f(\bar w^{s+1}_{\sv},\bar w^{s}_{\siii})}
 \frac{ K(\bar w^{s+1}_{\sv}| \bar w^{s}_{\sv})}{f(\bar w^{s+1}_{\sv}, \bar w^{s}_{\sv})}
 \frac{K(\bar w^{s+1}_{\siii}| \bar w^{s}_{\siii})}{f(\bar w^{s+1}_{\siii}, \bar w^{s}_{\siii})}=
 \frac{K(\bar w^{s+1}_{\sth}|\bar w^{s}_{\sth})}{f(\bar w^{s+1}_{\sth}|\bar w^{s}_{\sth})}.
 \end{equation}
This proves  \r{ac2} by induction over the cardinality $p$ of the set $\bar z$ in the case $i<j$.

 For $i\geq j$ the proof of \r{ac2} is similar.  \qed

\section{Eigenvector property of $\BB(\bar t)$}
\label{ApC}

Equation \r{ac2} yields the following formula for the action of the transfer matrix \r{trans} onto off-shell Bethe vector
$\BB(\bar t)$:
\begin{equation}\label{C1}
\mathfrak{t}(z) \mathbb{B}(\bar t) =  \lambda_{N+1}(z)
\sum_{i=1}^{N+1}      \sum_{{\rm part}} \mathbb{B}(\bar w_{\st} )
        \prod_{s=1}^{i-1} \frac{ f(\bar w^s_{\so},\bar w^s_{\st})}
        {h(\bar w^s_{\so}, \bar w^{s-1}_{\so}) f(\bar w^s_{\so},\bar w^{s-1}_{\st})}
		\prod_{s=i}^{N}      \frac{\alpha_{s}(\bar w^{s}_{\sth})   f(\bar w^s_{\st},\bar  w^s_{\sth})   }
		{h(\bar w^{s+1}_{\sth}, \bar w^{s}_{\sth})f(\bar w^{s+1}_{\st}, \bar w^{s}_{\sth})}.
\end{equation}
Here sum runs over partitions $\{\bar w^s_{\so},\bar w^s_{\st},\bar w^s_{\sth}\}\vdash \bar w^s=\{\bar t^s,z\}$
described in proposition~\ref{prop1}.

Let us select wanted terms from the action formula \r{C1} which
correspond to the partitions $\bar w^s_{\so}=z$ and $\bar w^s_{\st}=\bar t^s$ for $s=1,\ldots,i-1$ and
$\bar w^s_{\sth}=z$ and $\bar w^s_{\st}=\bar t^s$ for $s=i,\ldots,N+1$. Using
the facts that $h(z,z)=1$ and
\begin{equation*}
  \prod_{s=1}^{i-1} \frac{ f(z,\bar t^s)}{ f(z,\bar t^{s-1})}=f(z,\bar t^{i-1})\,,\quad\quad
  \prod_{s=i}^{N}      \frac{\alpha_{s}(z)   f(\bar t^s,z) }{f(\bar t^{s+1}, z)} =\frac{\lambda_i(z)f(\bar t^i,z)}{\lambda_{N+1}(z)},
\end{equation*}
we prove that wanted terms yield the right hand side of equation \r{BVcon} with
eigenvalue given by \r{BVeig}.

To prove that all other unwanted terms cancel each other provided the Bethe equations \r{BE}
are fulfilled, we consider the terms from the action of the diagonal monodromy matrix element
$T_{i+1,i+1}(z)$ which corresponds to the partitions
\begin{equation*}
\begin{split}
&\bar w^s_{\so}=z,\quad \bar w^s_{\st}=\bar t^s,\quad \bar w^s_{\sth}=\varnothing\quad\mbox{for}\quad s<i,\\
&\bar w^i_{\so}=t^i_\ell,\quad \bar w^i_{\st}=\{\bar t^i_\ell,z\},\quad \bar w^i_{\sth}=\varnothing\quad\mbox{for}\quad s=i,\\
&\bar w^s_{\so}=\varnothing,\quad \bar w^s_{\st}=\bar t^s,\quad \bar w^s_{\sth}=z\quad\mbox{for}\quad s>i.
\end{split}
\end{equation*}
We also consider the terms from the action of $T_{i,i}(z)$ corresponding to the partitions
\begin{equation*}
\begin{split}
&\bar w^s_{\so}=z,\quad \bar w^s_{\st}=\bar t^s,\quad \bar w^s_{\sth}=\varnothing\quad\mbox{for}\quad s<i,\\
&\bar w^i_{\so}=\varnothing,\quad \bar w^i_{\st}=\{\bar t^i_\ell,z\},\quad \bar w^i_{\sth}=t^i_\ell\quad\mbox{for}\quad s=i.\\
&\bar w^s_{\so}=\varnothing,\quad \bar w^s_{\st}=\bar t^s,\quad \bar w^s_{\sth}=z\quad\mbox{for}\quad s>i.
\end{split}
\end{equation*}
The terms from the right hand side of \r{C1} corresponding to both of these partitions can be written
as
\begin{equation*}
\begin{split}
\sum_{i=1}^{N+1}\sum_{\ell=1}^{r_i} \
&\BB(\bar t^1,\ldots,\bar t^{i-1},\{\bar t^i_{\ell},z\},\bar t^{i+1},\ldots, \bar t^{N})
\lambda_{i+1}(z)g(z,t^i_\ell) f(z,\bar t^{i-1})f(\bar t^{i+1},z)\\
&\quad \times 
\left[ \frac{\alpha_i(t^i_\ell)f(\bar t^i_\ell,t^i_\ell)}{f(\bar t^{i+1},t^i_\ell)}-
 \frac{f(t^i_\ell,\bar t^i_\ell)}{f(t^i_\ell,\bar t^{i-1})}\right] . 
\end{split}
\end{equation*}
 We see that these contributions disappear when the Bethe equations \r{BE} are satisfied.

In exactly the same way one can verify that all other unwanted terms disappear provided the Bethe equations
are satisfied.


\begin{thebibliography}{99}

\bibitem{HLPRS17} A.~A.~Hutsalyuk,  A.~Liashyk, S.~Z.~Pakuliak, E.~Ragoucy, N.~A.~Slavnov,
\textsl{Current presentation for the double super-Yangian $DY(\mathfrak{gl}(m|n))$ and Bethe vectors},
Russ. Math. Surv. {\bf 72}:1  (2017) 33--99 (Engl. transl.), \texttt{arXiv:1611.09620}, \texttt{doi: 10.1070/RM9754}.

\bibitem{EKhP07} B.~Enriquez, S.~Khoroshkin, S.~Pakuliak, \textsl{Weight
functions and Drinfeld currents,}
{Comm. Math. Phys.} {\bf 276} (2007) 691--725.

\bibitem{KhP-Kyoto} S. Khoroshkin, S. Pakuliak, {\sl A computation of an universal weight function for
the quantum affine algebra $U_q(\mathfrak{gl}(N))$},  {J. of Mathematics of Kyoto University},
{\bf 48} n.2 (2008) 277--321.

\bibitem{D88} V.~G.~Drinfeld. \textsl{A new realization of Yangians and of quantum affine algebras}, Soviet Math. Dokl. {\bf 36}
(1988) 212--216.

\bibitem{DF93} J.~Ding, I.~Frenkel. \textsl{Isomorphism of two realizations of quantum affine algebra
$U_q(\mathfrak{gl}(n))$}, Comm. Math. Phys. {\bf 156} (1993), 277–300.



\bibitem{JLM18} N.~Jing, M.~Liu, A.~Molev. \textsl{Isomorphism between the $R$-matrix and Drinfeld presentations
of Yangian in types $B$, $C$ and $D$}, Comm. Math. Phys. {\bf 361} (2018) 827--872.

\bibitem{Mol07} A. Molev,  \textsl{Yangians and Classical Lie Algebras}.
Mathematical Surveys and Monographs, 143. American Mathematical Society, Providence, RI, 2007.

\bibitem{KulRes81}
P. P. Kulish, N. Yu. Reshetikhin,
\textsl{Generalized Heisenberg ferromagnet and the Gross--Neveu model}, Zh. Eksp. Theor. Fiz.
{\bf 80} (1981) 214--228; Sov. Phys. JETP,  {\bf 53}:1 (1981)  108--114 (Engl. transl.)

\bibitem{KulRes83}
P. P. Kulish, N. Yu. Reshetikhin,
\textsl{Diagonalization of $GL(N)$ invariant transfer matrices and quantum $N$-wave system (Lee model)},
J.~Phys.~A:  {\bf 16} (1983) L591--L596.









\bibitem{LPRS-19}
A. Liashyk, S. Z. Pakuliak, E. Ragoucy, N. A. Slavnov.
{\sl Bethe vectors for orthogonal integrable models.}
Theoret. and Math. Phys. 201 (2019) 1545--1564.


\bibitem{GR-19}
A. Gerrard, V. Regelskis,
{\sl Nested algebraic Bethe ansatz for deformed orthogonal and symplectic spin chains},
{\tt arXiv:1912.11497}.

\bibitem{HLPRS17a} A.~A.~Hutsalyuk,  A.~Liashyk, S.~Z.~Pakuliak, E.~Ragoucy, N.~A.~Slavnov,
\textsl{Scalar products of Bethe vectors in the models with $\mathfrak{gl}(m|n)$ symmetry},
Nucl. Phys. B {\bf 923} (2017) 277--311, \texttt{arXiv:1704.08173}.

\bibitem{Kor82} V. E. Korepin, \textsl{Calculation of norms of Bethe wave functions}, Comm. Math. Phys. {\bf 86} (1982) 391--418.
%
\bibitem{Kor82a}
V. E.  Korepin, \textsl{Analysis of a bilinear relation for the six-vertex model},
Sov. Phys. Dokl., {\bf27} (1982) 612--613 (Engl. transl.).
%
\bibitem{IzeK83}
A. G. Izergin, V. E.  Korepin, \textsl{The problem of description of all $L$-operators for $R$-matrices of the models $XXX$ and $XXZ$} (in Russian),
Zap. Nauchn. Sem. LOMI, {\bf131} (1983) 80--87.
%
\bibitem{BogIK93L} V. E. Korepin, N. M. Bogoliubov,
A. G. Izergin, \textsl{Quantum Inverse Scattering Method and Correlation Functions}, Cambridge: Cambridge Univ.
Press, 1993.
%
\bibitem{Res86}
N. Yu. Reshetikhin, {\sl Calculation of the norm of Bethe vectors in models with $SU(3)$-symmetry},
Zap. Nauchn. Sem. LOMI {\bf 150} (1986) 196--213;    J. Math. Sci. {\bf 46} (1989) 1694--1706 (Engl. transl.).
%
\bibitem{HLPRS17b} A.~A.~Hutsalyuk,  A.~Liashyk, S.~Z.~Pakuliak, E.~Ragoucy, N.~A.~Slavnov,
\textsl{Norm of Bethe vectors in models with $\mathfrak{gl}(m|n)$ symmetry},
Nucl. Phys. B {\bf 926} (2018) 256--278, \texttt{arXiv:1705.09219}.
%
\bibitem{PakRS17} S. Pakuliak, E.  Ragoucy, and N. A. Slavnov,
{\sl Bethe vectors for models based on the super-Yangian $Y(\mathfrak{gl}(m|n))$},
J. Integrable Systems {\bf 2} (2017)  1--31, \texttt{arXiv:1604.02311}.

\bibitem{LPRS19} A.~Liashyk, S.~Z.~Pakuliak, E.~Ragoucy, N.~A.~Slavnov, \textsl{New symmetries of $\mathfrak{gl}(N)$-invariant
Bethe vectors}, J. Stat. Mech. Theory Exp., (2019) 044001.

\bibitem{BelPRS12a} S. Belliard, S. Pakuliak, E. Ragoucy, N. A. Slavnov, \textsl{The algebraic Bethe ansatz for scalar products in $SU(3)$-invariant integrable models},
J. Stat. Mech. Theory Exp., (2012) P10017, arXiv: \texttt{1207.0956}.

\bibitem{HLPRS18a} A.~A.~Hutsalyuk,  A.~Liashyk, S.~Z.~Pakuliak, E.~Ragoucy, N.~A.~Slavnov,
\textsl{Scalar products and norm of Bethe vectors for integrable models based on $U_q(\widehat{\mathfrak{gl}}(n))$},
SciPost Phys. {\bf 4} (2018) 006, \texttt{arXiv:1711.03867}.


\end{thebibliography}
\end{document}